\documentclass[10pt, letterpaper, reprint, superscriptaddress, aps, prl]
{revtex4-1} \usepackage[latin1]{inputenc} 
\usepackage{amsmath}
\usepackage{amsfonts} \usepackage{amssymb} \usepackage{float}
\usepackage{graphicx} \usepackage{soul} \usepackage[mathscr]{eucal}

\begin{document}

% **************************************************
% ******************** TITLE SECTION ***************
% **************************************************
\title{Dark Matter Detection Using Helium Evaporation and Field Ionization}
\author{Humphrey J. Maris} 
	\affiliation{Department of Physics, Brown University, Providence, Rhode Island 02912, USA} 
\author{George M. Seidel} 
	\affiliation{Department of Physics, Brown University, Providence, Rhode Island 02912, USA} 
\author{Derek Stein}
\affiliation{Department of Physics, Brown University, Providence, Rhode Island 02912, USA}

% **************************************************
% ******************** ABSTRACT ********************
% **************************************************

\begin{abstract} 
We describe a method for dark matter detection based on the evaporation of helium atoms from a cold surface and their subsequent detection using field ionization.  When a dark matter particle scatters off a nucleus of the target material, elementary excitations (phonons or rotons) are produced. 
Excitations which have an energy greater than the binding energy of helium to the surface can result in the evaporation of helium atoms. 
We propose to detect these atoms by ionizing them in a strong electric field. 
Because the binding energy of helium to surfaces can be below 1~meV, this detection scheme opens up new possibilities for the detection of dark matter particles in a mass range down to 1~MeV/c$^{2}$.

\end{abstract} 
\maketitle

Although the evidence for existence of dark matter from its gravitational interaction with ordinary matter is strong, direct detection of dark matter particles has not yet been achieved \cite{schneck2015,Akerib2017}. 
For a current plot of the experimentally excluded region of the scattering cross-section as a function of mass, see ref.~\cite{Akerib2017}. 
Recently, a number of theoretical models have been proposed in which the mass $m_\chi$  of dark matter particles would be below $\sim 10$~GeV/c$^2$, and thus below the mass range that can be easily detected in most of the current experiments \cite{Cushman2013}. 
In the search for low mass dark matter particles by direct detection, the energy threshold of the detector is the key parameter \cite{Strauss2016,Angloher2016}.
Present techniques based on electronic excitations of semiconductors \cite{Agnese2014,Agnese2015,Armengaud2016,Hehn2016}, scintillation from transparent crystals \cite{Strauss2016,Angloher2016}, and/or the thermal response of the target by sensitive thermometers \cite{Agnese2014,Hehn2016,Armengaud2016,Strauss2016,Angloher2016} all require the deposition of about 1 eV or more. 
Various techniques such as the use of narrow gap semiconductors \cite{Graham2012,Essig2012} or superconductors \cite{Hochberg2016} have been suggested to lower this energy threshold.
Here we describe an effective means of measuring energy depositions extending down to 1 meV based on the use of field ionization detection of helium atoms evaporated from surfaces at low temperatures. 

When a particle collides with a nucleus of mass $m_N$, the maximum transferable kinetic energy $T_{NR}$ is
\begin{equation}
T_{NR}\leq\frac{2m_\chi^2 m_N v_\chi^2}{(m_\chi + m_N)^2},
\label{eq:TNR}
\end{equation}
where $v_\chi$ is the relative speed of the particles. 
Recently there has been renewed interest in using liquid helium as a detector of light dark matter particles \cite{Guo2013,Ito2013,Schutz2016}, primarily because it is light and according to eq.~\ref{eq:TNR} receives more energy from a collision than would a higher mass target for $m_{\chi}<m_N$.
The high purity of liquid helium is also beneficial because it precludes internally generated background events due to radioactivity. 
Most significant for the purpose of increasing the sensitivity of a detector is the low binding energy of a helium atom to the liquid (0.62 meV) which must be overcome to release a helium atom into the vapor phase. 
The evaporation process provides a way to transfer, with amplification, energy deposited in the bulk of a target with a very large heat capacity to a calorimeter with much lower heat capacity, thereby lowering the minimum detectable $T_{NR}$ by several orders of magnitude relative to what can be achieved without its use.

%This favors the use of a helium-4 detector; helium-3 does have advantages \cite{Bradley1995} but for a full-scale detector this target is ruled out because of cost. 
%When a scattering event produces phonons and rotons, the quasiparticle excitations of the liquid, with energies greater than 0.62 meV, they are capable, upon hitting the free surface, of ejecting a helium from the liquid in a one to one process called quantum evaporation \cite{Brown1990,Wyatt1992}. 

\begin{figure}[!h]
\centering\includegraphics[width=\columnwidth]{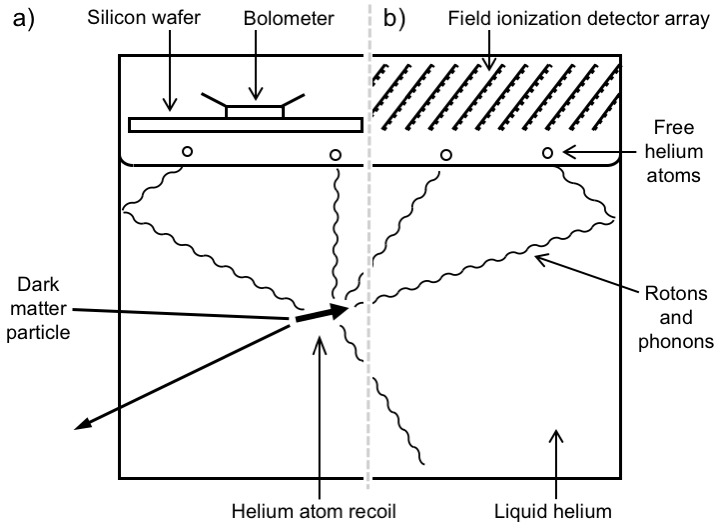}
\caption{Schematic diagram of (a) the original HERON experiment and (b) a dark matter direct detection experiment based on the detection of evaporated helium atoms by field ionization.  In both cases, the recoil of a helium nucleus in the superfluid produces rotons and phonons which, upon arriving at the free surface, cause helium atoms to be released by quantum evaporation.} 
\label{fig:schematic}
\end{figure}

A $^4$He-based detector for solar neutrinos (HERON) that made use of quantum evaporation of helium was proposed \cite{Lanou1987} and a series of experiments was performed to study the physics involved in the detection process \cite{Bandler1992,Bandler1995}.
Helium atoms that evaporated from superfluid helium as a result of charged particles stopped in the liquid \cite{Lanou1987,Bandler1992,Enss1994,Bandler1995,Adams1998,Adams2000} were detected by measuring the energy they imparted to a silicon wafer/calorimeter above the liquid helium.
A schematic diagram of the HERON detector is shown in Fig.~\ref{fig:schematic}a.
When the temperature is below 100~mK, the equilibrium density of $^4$He in the vapor phase is below $10^{-12}$~cm$^{-3}$ and the number of thermal rotons in the liquid is negligible.
A helium recoil results in a complex string of processes, but the final outcome is the production of phonon and roton excitations \cite{Maris1977}.
The phonons and the rotons that have an energy above about 0.7 meV propagate through the liquid without scattering or decay \cite{NoteLosses}.
When one of these excitations reaches the free surface of the helium, a process called quantum evaporation can eject a helium atom \cite{Brown1990,Wyatt1992,Balibar2016}.
The wafer/calorimeter was positioned above the liquid surface in a way that maintained the wafer free of a superfluid helium film \cite{Torii1992}.%, which is important for two reasons: 
%First, at temperatures below 100~mK the heat capacity of a saturated helium film is much larger than the heat capacity of the silicon wafer. 
%Second, when a helium atom is adsorbed onto a bare surface of silicon the heat of adsorption per atom is more than 10 times larger than the binding energy to the liquid, providing a very beneficial amplification of the energy and the measured temperature rise of the wafer. 

The calorimetric detection of evaporated helium atoms as employed by HERON is a powerful method for decreasing the threshold energy measurable in a large-target-mass detector. 
When applied to the search for dark matter, the lowest detectable $m_\chi$ is determined by the minimum number of evaporated helium atoms that can be detected. 
Single-atom sensitivity would enable the detection of WIMPs as light as 0.6~MeV/c$^2$, according to eq.~\ref{eq:TNR} and assuming $v_\chi=537$~km/s, the galactic escape velocity in our region. 
At present, no large-area calorimeter has demonstrated an energy threshold close to being able to sense a single helium atom.

\begin{figure}[!h]
\centering\includegraphics[width=\columnwidth]{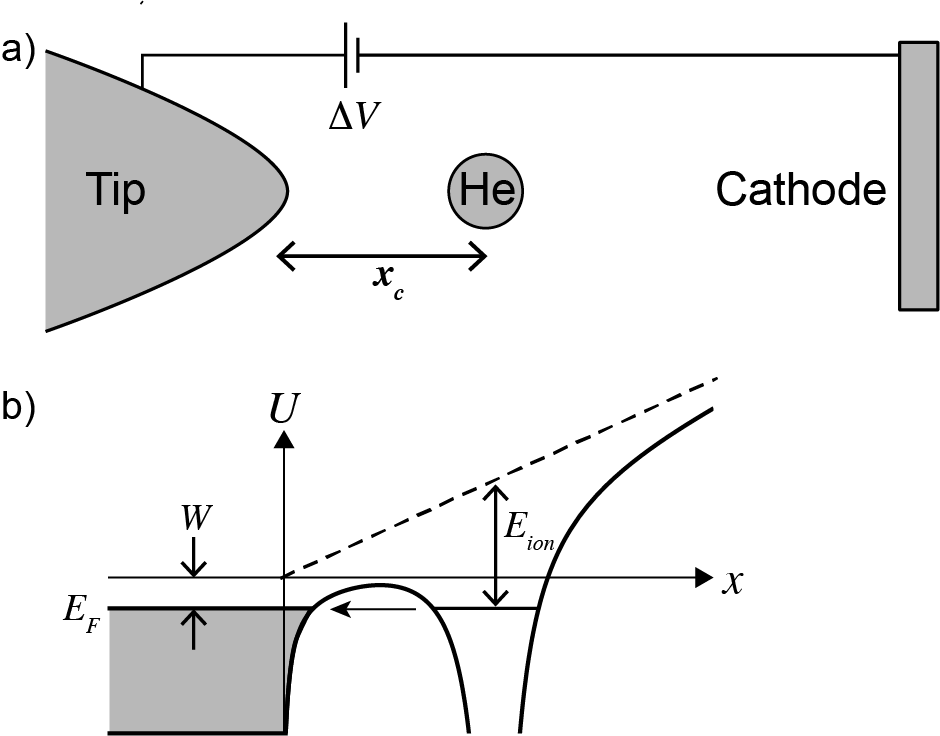}
\caption{Field ionization. a) Illustration of a helium atom a distance $x_c$ away from a positively charged metal tip. b) Diagram of the electron potential energy $U$ showing the field-distorted barrier of a bound electron (black line), the influence of the applied field on a free electron (dashed line), the work function of the tip $W$, and the helium ionization energy $E_{ion}$.} 
\label{fig:FImechanism}
\end{figure}

Field ionization offers a method for detecting single free helium atoms. 
The high electric fields in the vicinity of a sharp, positively charged, metal tip can ionize helium atoms, whereupon the resulting positive ion accelerates and impinges on a cathode with a high kinetic energy. 
A calorimeter can easily measure the impact of even a single ion on the cathode. 
Field ionization is a well-studied process, having been discovered in 1951 by M{\"u}ller \cite{Muller1951} and used extensively in field ion microscopy, which M{\"u}ller invented \cite{Muller1951,Muller1956,Gomer1994}.
It involves the tunneling of an electron from a neutral atom to a charged metal tip through a field-distorted barrier, illustrated schematically in Fig.~\ref{fig:FImechanism}.
The tip is held at a positive voltage $\Delta V$, typically several kV, relative to a distant cathode.
An electron can tunnel from the atom to the tip when the electric field raises the electron's energy above the Fermi level of the tip. 
The critical distance $x_c$ from the tip beyond which that condition is met decreases with increasing field strength $E$.
The tunneling probability increases rapidly with $E$ because the width of the tunneling barrier shrinks. 
M{\"u}ller found that field ionization of gas phase helium atoms by a single tip produced a measurably large electrical current when $E\approx 2$ V/\AA, which corresponds to $x_c$ of the order of a few \AA ~\cite{Muller1956}.

The efficiency with which a field ionization detector detects evaporated helium atoms will depend on several factors, including the electric field at the tip, the field gradients around the tip, the tip surface, the temperature, and the geometry of the detector. 
The field near the tip controls the tunneling process. 
The probability that a helium atom approaching the tip will become ionized before it reaches $x_c$ is essentially unity when $E=5$ V/\AA, because the time the atom takes to pass through the high field region is longer than the inverse tunneling rate \cite{ODonnell2010}. 
Furthermore, the ions consistently form at a position where the potential is only 10 to 20 volts below $\Delta V$ because ionization occurs overwhelmingly within a narrow region extending less than 1 \AA~ beyond $x_c$ \cite{Muller1956,ODonnell2010}; this ensures that each ion will deposit a large and consistent amount of kinetic energy in a calorimeter located at the cathode.

To quantify the atom gathering power of a field emission tip, one commonly defines an effective capture radius $R_{eff}$ such that the current $I$ of the emitted ions is equal to $\pi R_{eff}Je$, where $-e$ is the electron charge and $J$ is the flux of helium atoms onto the surface. 
At pressure $P$ and temperature $T$, $J=P/\sqrt{2\pi m_N k_BT}$, where $m_N$ is the mass of a helium atom. 
When $\Delta V$ is high, $R_{eff}$ can significantly exceed the nominal tip radius. 
For example, early measurements by M{\"u}ller and Bahadur \cite{Muller1956} and by Johnston and King \cite{Johnston1966} obtained $R_{eff}$ as high as 450~nm and 500~nm, respectively, using similar tungsten tips with nominal radii of 100~nm.

We expect that helium atoms can be efficiently field ionized below 100~mK because large values of $R_{eff}$ are the result of several effects which can be enhanced by design. 
The inhomogeneous fields around a charged tip exert a polarization force on helium atoms that gathers them inward. 
The potential from the polarization interaction is $U_{pol}=-\frac{1}{2}\alpha E^2=-7.1E^2$~meV, where $E$ is measured in V/\AA~ and $\alpha$ is the polarizability of atomic helium.
The range over which $U_{pol}$ collects atoms grows as the kinetic energy of the atoms decreases. 
The atoms evaporated from helium by rotons have a range of energy from 0.13~meV up to about 0.7~meV, and thus their trajectory will be strongly influenced by $U_{pol}$. 
The effect of $U_{pol}$ is further assisted by the small fraction of kinetic energy that helium atoms lose upon hitting the surface near the tip \cite{Goodman1980}, which allows them to bounce off the surface many times before becoming adsorbed, and as they bounce the polarization force can guide them toward the tip \cite{ODonnell2010}. 
Accordingly, the helium field ionization current at a tungsten tip increases rapidly as the temperature decreases \cite{Southon1963,Borret1990,Piskur2008}.
Finally, we note that adsorbed atoms may have a significant mobility after adsorption. 
Halpern and Gomer \cite{Halpern1969} have shown that helium migrates easily towards a charged tip with helium films on the surface of tungsten at 4.2~K. It is possible that as a result of quantum tunneling there is still significant diffusion even at $T=0$~K. 

\begin{figure}[!ht]
\centering\includegraphics[width=\columnwidth]{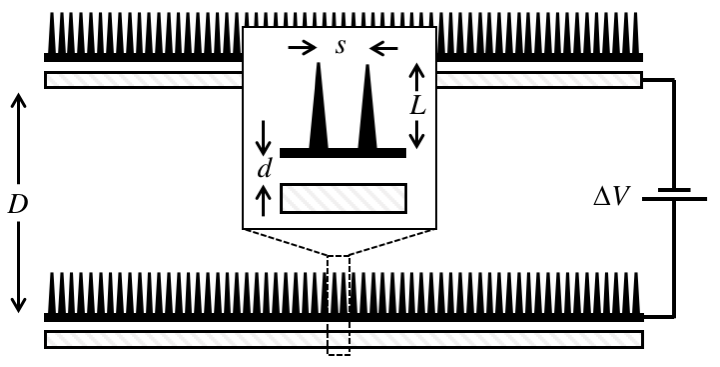}
\caption{Schematic diagram of a field ionization tip array indicating $D$, $L$, $s$, and $d$.} 
\label{fig:FITipArray}
\end{figure}

A helium atom detector for dark matter searches could be configured as an array of tips on the anode in a parallel-plate capacitor geometry, as sketched in Fig.~\ref{fig:FITipArray}. 
The cathode could be a thin wafer with a metalized surface connected to a calorimeter. 
A dense array of tips should maximize the probability that approaching helium atoms will become captured and ionized, while arranging many inclined arrays in rows as sketched in Fig.~\ref{fig:schematic}b should result in a low fraction of free helium atoms that reflect off a detector surface back toward the target.  
The important dimensions of the detector geometry include the tip radius $R$, the length $L$ of each tip, the lateral spacing $s$ between tips in the array, the spacing $D$ between the anode and the cathode, and the thickness $d$ of the gap between neighboring arrays. 
Sharp tips achieve the critical field to induce ionization by locally enhancing the macroscopic field, given by $E_m=\Delta V/D$, by a factor $\gamma$. 
For a single tip shaped like a cylindrical post with a hemispherical cap, $\gamma$ depends on the ratio $L/R$ as $\gamma=1.125(L/R+2)^{0.91}$ \cite{Edgcombe2001}. 
A tip with $R=10$~nm and $L=20$~$\mu$m thus achieves an enhancement factor $\gamma = 1140$. 
In an array, each tip is screened by its neighbors, resulting in a decrease in $\gamma$, but shielding only becomes significant when $s$ is small compared with $L$. 
Calculations by Read and Bowring \cite{Read2004} found that $\gamma$ is reduced by only about 10 \% relative to an isolated tip when $s/L=0.1$, and so the tip considered above could be arrayed with $s=2 ~\mu$m and still achieve a field enhancement factor of 1000. 
To generate  $E=5$~V/\AA~ at the tips with $\Delta V=50$~kV, $D$ could be as wide as 1~mm. 
%The dielectric strength of silicon dioxide is $10^9$~V/m, so neighboring arrays operating at $\Delta V=50$~kV can be isolated from each other with a thin dielectric spacer of thickness $d\sim 50$~$\mu$m.
A thin ($d<D$) spacer with a high dielectric strength can isolate neighboring arrays operating at $\Delta V=50$~kV.
Arrays of tips with similar dimensions to those described above have been fabricated by scalable wafer processing techniques and used for field ionization applications \cite{Johnson2012}.

While tungsten tips, as discussed above, have been used extensively to study the physics of field ionization, more recently large arrays of carbon nanotubes \cite{Modi2003}, and metallized semiconducting tips \cite{Gesemann2011,Liu2013,Spitsina2016}, have been fabricated for a variety of applications of field ionization. 
Many designs with anode to cathode separations in the micrometer range can produce ionization at applied potentials below 100~V \cite{Mohammadpour2014}. 
In the case of semiconducting tips, the combination of field penetration into the semiconductor \cite{Ohno1978,Tsong1979} and band bending at the surface and empty acceptor states on the surface of p-type materials \cite{Tsong1979} result in field ionization of helium with potentials below 10 kV applied across a 2 mm anode to cathode gap. 
These new materials and configurations open up many different possibilities for the design of a low temperature sensor for evaporated helium. 
Studies of sensitivity, operating potential, and difficulty of fabrication are required to assess the best approach to a design.

The choice and processing of materials used in the field ionization detector are important because they influence a number of effects that could adversely influence the accurate characterization of evaporated atoms. 
When a helium ion strikes the cathode of the detector with several keV of kinetic energy, atoms can be sputtered from the metal and, independently, secondary electrons can be emitted. 
The probability of sputtering is small and its occurrence can be identified and rendered inconsequential, and while the probability of secondary electron emission can be much larger, the end result of such emission is not serious.
Dark counts are a separate issue that must be addressed in single particle detectors. 
A field ionization detector for helium atoms has two dark count mechanisms that are unique to its design and operation, in addition to the obvious possibility of field emission of electrons from the cathode \cite{Suzuki2001,Lepetit2016}. 
One process occurs if the field at the tips becomes too high and atoms composing the tip become detached and ionized, a process called field evaporation. 
The other mechanism involves helium atoms adsorbed on the shank possibly migrating to the high field tip and field evaporating.
We believe it is possible to identify and mitigate these adverse effects so as to render them unimportant. 
A detailed discussion of these various adverse effects, the probability of their occurrence, and methods to mitigate them, is contained in the Supplemental Material.

The probability that elementary excitations generated in the target by a nuclear recoil will result in the emission of a helium atom from the surface is an important characteristic of the detector.
Although it does not affect the threshold energy sensitivity, it does affect the overall recoil detection efficiency and the energy resolution.
In superfluid helium, kinematic constraints limit the incident angle of quasiparticles to the surface for which quantum evaporation is allowed. Evaporation by the predominant roton excitations is constrained to a narrow cone of approximately 25 degrees with respect to the vertical \cite{Enss1994}. As a consequence, only about 5 \% of the quasiparticles generated by a nuclear recoil will, without reflection from walls, be capable of evaporating an atom.
Of those rotons that fall within the allowed cone it is estimated, based on data in ref. \cite{BandlerPhD1995}, that 50 \% lead to evaporation, for an overall probability of evaporation of 2.5 \%. 
This is in rough accord with theory \cite{Guilleumas1998}.
Rotons that reflect from walls of the liquid container may eventually arrive at the free surface at an angle able to evaporate atoms. 
However, because of the large number of reflections that on average are required for this to occur, even low probability loss mechanisms at walls make unreliable any estimation of the fraction which induce evaporation. 
Typically, the energy transmission through helium/solid interfaces is found experimentally \cite{Wyatt1976} to be considerably larger for excitations of order of a meV than is calculated theoretically \cite{Khalatnikov1965}. 
%While this is detrimental to the performance of a superfluid helium detector, it is beneficial to one based on the evaporation of a helium film on a solid.

The detection of evaporated helium atoms resulting from the recoil of nuclei in a target is not limited to WIMP scattering in a bath of superfluid helium. 
Atoms can also be evaporated from helium film-coated surfaces of crystalline solids in which the phonon mean free path is long.  
Cryogenic dark matter detectors currently being used to search for WIMPs of mass of ~10 GeV/c$^2$ \cite{schneck2015} employ ultra-pure crystals with low phonon scattering. 
The detection of evaporated helium from the surfaces of these or similar crystals could lower the energy threshold and hence the measureable WIMP mass of these detectors by orders of magnitude. 
%The dark matter detection strategy described here can be implemented just as well using a crystalline solid target with long phonon mean free paths, such as the germanium crystals used in ongoing dark matter searches \cite{schneck2015}. 
In this regard, Sinvani {\it et al} \cite{Sinvani1983} observed the evaporation of helium from a superfluid film on a 1 cm sapphire crystal to which they applied a short heat pulse. 
From timing discrimination, they identified evaporation associated with both longitudinal and transverse phonons in the crystal. 
More {\it et al} \cite{More1996} demonstrated evaporation of helium atoms from a silicon wafer covered on one side with a $\sim 20$~nm thick saturated superfluid helium film, but bare on the other side. 
When an alpha particle was directed at the bare side the efficiency of conversion of energy deposition in silicon to evaporated atoms was comparable to that found for bulk helium. 
Similar results were found for a 0.1 cm thick NaF crystal \cite{More1996}, but no evaporation was detected from a glass microscope slide, as expected because of the absence of ballistic phonon propagation. 
%The conversion of energy within the phonon system in the solid to evaporated helium atoms most likely depends on surface properties and thickness of the superfluid film. 
%While the energy per helium atom on the first adsorption layer is typically in the range of 3 to 6 meV, the binding energy rapidly drops for the second and additional layers of helium. 
%The optimal film thickness for energy transfer from phonons in the bulk solid to evaporated atoms remains to be determined. 
A potential advantage of using a solid target covered with a thin layer of adsorbed helium is the possibility of defining a fiducial volume within the target and discriminating background signals based on the distribution of helium atoms detected following an event. 
Finally, a solid crystalline target offers a means to further reduce the threshold phonon energy required to evaporate helium atoms. 
$^4$He is very weakly bound to the alkali metals, with an adsorption energy ranging from 1.2 meV for lithium \cite{Van2008} to 0.33 meV for cesium \cite{Rutledge1992,Taborek1992,Gatica2009}. 
The binding energy for $^3$He is even less, being only 0.17 meV on cesium \cite{Ross1995}. 
If a film of Cs no more than a few monolayers thick were deposited on a target crystal, followed by a monolayer of helium, the likely consequence is that energy within the phonon system would be transferred to the surface and lead to helium evaporation. 
The deposition of the reactive alkalis adds complexity to any low temperature experiment, but it has been successfully demonstrated in a variety of different circumstances \cite{Van2008,Ross1995,Nacher1991}.

We currently lack the experimental knowledge needed to better estimate a number of other important detector characteristics, such as the sensitivity, energy resolution, time resolution, background sources and rejection, and dark counts. 
Further studies of the physics in two areas seem particularly important to us: First, the processes that fundamental excitations of the target undergo at the boundaries, including possibly reflection, down-conversion to lower energy excitations, production of surface riplons, or release of free helium atoms, will strongly influence the relationship between the energy deposited and the number of free helium atoms released, but the probability of their occurrence is not known. 
Second, the migration of helium atoms along the surfaces of the field ionization detector, which will influence the detection efficiency and time resolution of the detector, and possibly give rise to dark counts, needs to be better understood.

In summary, quantum evaporation of helium atoms from cold surfaces and their subsequent detection by field ionization provides a powerful technique for measuring energy deposits of less than 1 meV in a large target mass. 
The wide range of suitable target materials includes superfluid helium and many crystalline solids possessing long phonon mean free paths. 
The technique also conveniently separates the energy absorption process in the target from the process by which evaporated helium atoms are measured. 
Field ionization is an appealing process for measurement because it offers single-atom sensitivity and can be implemented using arrays of sharp tips that cover wide areas. 
When applied to the direct detection of dark matter, it opens up the possibility of searching for particles in the range of 1 MeV/c$^2$. 
Finally, we note that this technique may be applicable in other areas, for instance calorimetric experiments that require large masses like searches for neutrinoless double beta decay and the spectroscopy of photons over wide ranges of energy. 
%Further work is needed to experimentally realize the proposed detector and quantify the detection efficiency of helium atoms, the energy resolution, and the dark count rate. 

\begin{acknowledgments}
We thank Benjamin Wiener, Rick Gaitskell, Andrei Korotkov, David Cutts, Jiji Fan, Bob Lanou and the participants of the U.S. Cosmic Visions: New Ideas in Dark Matter workshop for stimulating discussions. This work was supported in part by the US National Science Foundation through Grant No. DMR-1505044. 
\end{acknowledgments}

\bibliography{references}

\begin{thebibliography}{61}%
\makeatletter
\providecommand \@ifxundefined [1]{%
 \@ifx{#1\undefined}
}%
\providecommand \@ifnum [1]{%
 \ifnum #1\expandafter \@firstoftwo
 \else \expandafter \@secondoftwo
 \fi
}%
\providecommand \@ifx [1]{%
 \ifx #1\expandafter \@firstoftwo
 \else \expandafter \@secondoftwo
 \fi
}%
\providecommand \natexlab [1]{#1}%
\providecommand \enquote  [1]{``#1''}%
\providecommand \bibnamefont  [1]{#1}%
\providecommand \bibfnamefont [1]{#1}%
\providecommand \citenamefont [1]{#1}%
\providecommand \href@noop [0]{\@secondoftwo}%
\providecommand \href [0]{\begingroup \@sanitize@url \@href}%
\providecommand \@href[1]{\@@startlink{#1}\@@href}%
\providecommand \@@href[1]{\endgroup#1\@@endlink}%
\providecommand \@sanitize@url [0]{\catcode `\\12\catcode `\$12\catcode
  `\&12\catcode `\#12\catcode `\^12\catcode `\_12\catcode `\%12\relax}%
\providecommand \@@startlink[1]{}%
\providecommand \@@endlink[0]{}%
\providecommand \url  [0]{\begingroup\@sanitize@url \@url }%
\providecommand \@url [1]{\endgroup\@href {#1}{\urlprefix }}%
\providecommand \urlprefix  [0]{URL }%
\providecommand \Eprint [0]{\href }%
\providecommand \doibase [0]{http://dx.doi.org/}%
\providecommand \selectlanguage [0]{\@gobble}%
\providecommand \bibinfo  [0]{\@secondoftwo}%
\providecommand \bibfield  [0]{\@secondoftwo}%
\providecommand \translation [1]{[#1]}%
\providecommand \BibitemOpen [0]{}%
\providecommand \bibitemStop [0]{}%
\providecommand \bibitemNoStop [0]{.\EOS\space}%
\providecommand \EOS [0]{\spacefactor3000\relax}%
\providecommand \BibitemShut  [1]{\csname bibitem#1\endcsname}%
\let\auto@bib@innerbib\@empty
%</preamble>
\bibitem [{\citenamefont {Schneck}\ \emph {et~al.}(2015)\citenamefont
  {Schneck}, \citenamefont {Cabrera}, \citenamefont {Cerde\~no}, \citenamefont
  {Mandic}, \citenamefont {Rogers}, \citenamefont {Agnese}, \citenamefont
  {Anderson}, \citenamefont {Asai}, \citenamefont {Balakishiyeva},
  \citenamefont {Barker}, \citenamefont {Basu~Thakur}, \citenamefont {Bauer},
  \citenamefont {Billard}, \citenamefont {Borgland}, \citenamefont {Brandt},
  \citenamefont {Brink}, \citenamefont {Bunker}, \citenamefont {Caldwell},
  \citenamefont {Calkins}, \citenamefont {Chagani}, \citenamefont {Chen},
  \citenamefont {Cooley}, \citenamefont {Cornell}, \citenamefont {Crewdson},
  \citenamefont {Cushman}, \citenamefont {Daal}, \citenamefont {Di~Stefano},
  \citenamefont {Doughty}, \citenamefont {Esteban}, \citenamefont {Fallows},
  \citenamefont {Figueroa-Feliciano}, \citenamefont {Godfrey}, \citenamefont
  {Golwala}, \citenamefont {Hall}, \citenamefont {Harris}, \citenamefont
  {Hofer}, \citenamefont {Holmgren}, \citenamefont {Hsu}, \citenamefont
  {Huber}, \citenamefont {Jardin}, \citenamefont {Jastram}, \citenamefont
  {Kamaev}, \citenamefont {Kara}, \citenamefont {Kelsey}, \citenamefont
  {Kennedy}, \citenamefont {Leder}, \citenamefont {Loer}, \citenamefont
  {Lopez~Asamar}, \citenamefont {Lukens}, \citenamefont {Mahapatra},
  \citenamefont {McCarthy}, \citenamefont {Mirabolfathi}, \citenamefont
  {Moffatt}, \citenamefont {Morales~Mendoza}, \citenamefont {Oser},
  \citenamefont {Page}, \citenamefont {Page}, \citenamefont {Partridge},
  \citenamefont {Pepin}, \citenamefont {Phipps}, \citenamefont {Prasad},
  \citenamefont {Pyle}, \citenamefont {Qiu}, \citenamefont {Rau}, \citenamefont
  {Redl}, \citenamefont {Reisetter}, \citenamefont {Ricci}, \citenamefont
  {Roberts}, \citenamefont {Saab}, \citenamefont {Sadoulet}, \citenamefont
  {Sander}, \citenamefont {Schnee}, \citenamefont {Scorza}, \citenamefont
  {Serfass}, \citenamefont {Shank}, \citenamefont {Speller}, \citenamefont
  {Toback}, \citenamefont {Upadhyayula}, \citenamefont {Villano}, \citenamefont
  {Welliver}, \citenamefont {Wilson}, \citenamefont {Wright}, \citenamefont
  {Yang}, \citenamefont {Yellin}, \citenamefont {Yen}, \citenamefont {Young},\
  and\ \citenamefont {Zhang}}]{schneck2015}%
  \BibitemOpen
  \bibfield  {author} {\bibinfo {author} {\bibfnamefont {K.}~\bibnamefont
  {Schneck}}, \bibinfo {author} {\bibfnamefont {B.}~\bibnamefont {Cabrera}},
  \bibinfo {author} {\bibfnamefont {D.~G.}\ \bibnamefont {Cerde\~no}}, \bibinfo
  {author} {\bibfnamefont {V.}~\bibnamefont {Mandic}}, \bibinfo {author}
  {\bibfnamefont {H.~E.}\ \bibnamefont {Rogers}}, \bibinfo {author}
  {\bibfnamefont {R.}~\bibnamefont {Agnese}}, \bibinfo {author} {\bibfnamefont
  {A.~J.}\ \bibnamefont {Anderson}}, \bibinfo {author} {\bibfnamefont
  {M.}~\bibnamefont {Asai}}, \bibinfo {author} {\bibfnamefont {D.}~\bibnamefont
  {Balakishiyeva}}, \bibinfo {author} {\bibfnamefont {D.}~\bibnamefont
  {Barker}}, \bibinfo {author} {\bibfnamefont {R.}~\bibnamefont {Basu~Thakur}},
  \bibinfo {author} {\bibfnamefont {D.~A.}\ \bibnamefont {Bauer}}, \bibinfo
  {author} {\bibfnamefont {J.}~\bibnamefont {Billard}}, \bibinfo {author}
  {\bibfnamefont {A.}~\bibnamefont {Borgland}}, \bibinfo {author}
  {\bibfnamefont {D.}~\bibnamefont {Brandt}}, \bibinfo {author} {\bibfnamefont
  {P.~L.}\ \bibnamefont {Brink}}, \bibinfo {author} {\bibfnamefont
  {R.}~\bibnamefont {Bunker}}, \bibinfo {author} {\bibfnamefont {D.~O.}\
  \bibnamefont {Caldwell}}, \bibinfo {author} {\bibfnamefont {R.}~\bibnamefont
  {Calkins}}, \bibinfo {author} {\bibfnamefont {H.}~\bibnamefont {Chagani}},
  \bibinfo {author} {\bibfnamefont {Y.}~\bibnamefont {Chen}}, \bibinfo {author}
  {\bibfnamefont {J.}~\bibnamefont {Cooley}}, \bibinfo {author} {\bibfnamefont
  {B.}~\bibnamefont {Cornell}}, \bibinfo {author} {\bibfnamefont {C.~H.}\
  \bibnamefont {Crewdson}}, \bibinfo {author} {\bibfnamefont {P.}~\bibnamefont
  {Cushman}}, \bibinfo {author} {\bibfnamefont {M.}~\bibnamefont {Daal}},
  \bibinfo {author} {\bibfnamefont {P.~C.~F.}\ \bibnamefont {Di~Stefano}},
  \bibinfo {author} {\bibfnamefont {T.}~\bibnamefont {Doughty}}, \bibinfo
  {author} {\bibfnamefont {L.}~\bibnamefont {Esteban}}, \bibinfo {author}
  {\bibfnamefont {S.}~\bibnamefont {Fallows}}, \bibinfo {author} {\bibfnamefont
  {E.}~\bibnamefont {Figueroa-Feliciano}}, \bibinfo {author} {\bibfnamefont
  {G.~L.}\ \bibnamefont {Godfrey}}, \bibinfo {author} {\bibfnamefont {S.~R.}\
  \bibnamefont {Golwala}}, \bibinfo {author} {\bibfnamefont {J.}~\bibnamefont
  {Hall}}, \bibinfo {author} {\bibfnamefont {H.~R.}\ \bibnamefont {Harris}},
  \bibinfo {author} {\bibfnamefont {T.}~\bibnamefont {Hofer}}, \bibinfo
  {author} {\bibfnamefont {D.}~\bibnamefont {Holmgren}}, \bibinfo {author}
  {\bibfnamefont {L.}~\bibnamefont {Hsu}}, \bibinfo {author} {\bibfnamefont
  {M.~E.}\ \bibnamefont {Huber}}, \bibinfo {author} {\bibfnamefont {D.~M.}\
  \bibnamefont {Jardin}}, \bibinfo {author} {\bibfnamefont {A.}~\bibnamefont
  {Jastram}}, \bibinfo {author} {\bibfnamefont {O.}~\bibnamefont {Kamaev}},
  \bibinfo {author} {\bibfnamefont {B.}~\bibnamefont {Kara}}, \bibinfo {author}
  {\bibfnamefont {M.~H.}\ \bibnamefont {Kelsey}}, \bibinfo {author}
  {\bibfnamefont {A.}~\bibnamefont {Kennedy}}, \bibinfo {author} {\bibfnamefont
  {A.}~\bibnamefont {Leder}}, \bibinfo {author} {\bibfnamefont
  {B.}~\bibnamefont {Loer}}, \bibinfo {author} {\bibfnamefont {E.}~\bibnamefont
  {Lopez~Asamar}}, \bibinfo {author} {\bibfnamefont {P.}~\bibnamefont
  {Lukens}}, \bibinfo {author} {\bibfnamefont {R.}~\bibnamefont {Mahapatra}},
  \bibinfo {author} {\bibfnamefont {K.~A.}\ \bibnamefont {McCarthy}}, \bibinfo
  {author} {\bibfnamefont {N.}~\bibnamefont {Mirabolfathi}}, \bibinfo {author}
  {\bibfnamefont {R.~A.}\ \bibnamefont {Moffatt}}, \bibinfo {author}
  {\bibfnamefont {J.~D.}\ \bibnamefont {Morales~Mendoza}}, \bibinfo {author}
  {\bibfnamefont {S.~M.}\ \bibnamefont {Oser}}, \bibinfo {author}
  {\bibfnamefont {K.}~\bibnamefont {Page}}, \bibinfo {author} {\bibfnamefont
  {W.~A.}\ \bibnamefont {Page}}, \bibinfo {author} {\bibfnamefont
  {R.}~\bibnamefont {Partridge}}, \bibinfo {author} {\bibfnamefont
  {M.}~\bibnamefont {Pepin}}, \bibinfo {author} {\bibfnamefont
  {A.}~\bibnamefont {Phipps}}, \bibinfo {author} {\bibfnamefont
  {K.}~\bibnamefont {Prasad}}, \bibinfo {author} {\bibfnamefont
  {M.}~\bibnamefont {Pyle}}, \bibinfo {author} {\bibfnamefont {H.}~\bibnamefont
  {Qiu}}, \bibinfo {author} {\bibfnamefont {W.}~\bibnamefont {Rau}}, \bibinfo
  {author} {\bibfnamefont {P.}~\bibnamefont {Redl}}, \bibinfo {author}
  {\bibfnamefont {A.}~\bibnamefont {Reisetter}}, \bibinfo {author}
  {\bibfnamefont {Y.}~\bibnamefont {Ricci}}, \bibinfo {author} {\bibfnamefont
  {A.}~\bibnamefont {Roberts}}, \bibinfo {author} {\bibfnamefont
  {T.}~\bibnamefont {Saab}}, \bibinfo {author} {\bibfnamefont {B.}~\bibnamefont
  {Sadoulet}}, \bibinfo {author} {\bibfnamefont {J.}~\bibnamefont {Sander}},
  \bibinfo {author} {\bibfnamefont {R.~W.}\ \bibnamefont {Schnee}}, \bibinfo
  {author} {\bibfnamefont {S.}~\bibnamefont {Scorza}}, \bibinfo {author}
  {\bibfnamefont {B.}~\bibnamefont {Serfass}}, \bibinfo {author} {\bibfnamefont
  {B.}~\bibnamefont {Shank}}, \bibinfo {author} {\bibfnamefont
  {D.}~\bibnamefont {Speller}}, \bibinfo {author} {\bibfnamefont
  {D.}~\bibnamefont {Toback}}, \bibinfo {author} {\bibfnamefont
  {S.}~\bibnamefont {Upadhyayula}}, \bibinfo {author} {\bibfnamefont {A.~N.}\
  \bibnamefont {Villano}}, \bibinfo {author} {\bibfnamefont {B.}~\bibnamefont
  {Welliver}}, \bibinfo {author} {\bibfnamefont {J.~S.}\ \bibnamefont
  {Wilson}}, \bibinfo {author} {\bibfnamefont {D.~H.}\ \bibnamefont {Wright}},
  \bibinfo {author} {\bibfnamefont {X.}~\bibnamefont {Yang}}, \bibinfo {author}
  {\bibfnamefont {S.}~\bibnamefont {Yellin}}, \bibinfo {author} {\bibfnamefont
  {J.~J.}\ \bibnamefont {Yen}}, \bibinfo {author} {\bibfnamefont {B.~A.}\
  \bibnamefont {Young}}, \ and\ \bibinfo {author} {\bibfnamefont
  {J.}~\bibnamefont {Zhang}} (\bibinfo {collaboration} {SuperCDMS
  Collaboration}),\ }\href {\doibase 10.1103/PhysRevD.91.092004} {\bibfield
  {journal} {\bibinfo  {journal} {Phys. Rev. D}\ }\textbf {\bibinfo {volume}
  {91}},\ \bibinfo {pages} {092004} (\bibinfo {year} {2015})}\BibitemShut
  {NoStop}%
\bibitem [{\citenamefont {Akerib}\ \emph {et~al.}(2017)\citenamefont {Akerib},
  \citenamefont {Alsum}, \citenamefont {Ara\'ujo}, \citenamefont {Bai},
  \citenamefont {Bailey}, \citenamefont {Balajthy}, \citenamefont {Beltrame},
  \citenamefont {Bernard}, \citenamefont {Bernstein}, \citenamefont
  {Biesiadzinski}, \citenamefont {Boulton}, \citenamefont {Bramante},
  \citenamefont {Br\'as}, \citenamefont {Byram}, \citenamefont {Cahn},
  \citenamefont {Carmona-Benitez}, \citenamefont {Chan}, \citenamefont
  {Chiller}, \citenamefont {Chiller}, \citenamefont {Currie}, \citenamefont
  {Cutter}, \citenamefont {Davison}, \citenamefont {Dobi}, \citenamefont
  {Dobson}, \citenamefont {Druszkiewicz}, \citenamefont {Edwards},
  \citenamefont {Faham}, \citenamefont {Fiorucci}, \citenamefont {Gaitskell},
  \citenamefont {Gehman}, \citenamefont {Ghag}, \citenamefont {Gibson},
  \citenamefont {Gilchriese}, \citenamefont {Hall}, \citenamefont {Hanhardt},
  \citenamefont {Haselschwardt}, \citenamefont {Hertel}, \citenamefont {Hogan},
  \citenamefont {Horn}, \citenamefont {Huang}, \citenamefont {Ignarra},
  \citenamefont {Ihm}, \citenamefont {Jacobsen}, \citenamefont {Ji},
  \citenamefont {Kamdin}, \citenamefont {Kazkaz}, \citenamefont {Khaitan},
  \citenamefont {Knoche}, \citenamefont {Larsen}, \citenamefont {Lee},
  \citenamefont {Lenardo}, \citenamefont {Lesko}, \citenamefont {Lindote},
  \citenamefont {Lopes}, \citenamefont {Manalaysay}, \citenamefont {Mannino},
  \citenamefont {Marzioni}, \citenamefont {McKinsey}, \citenamefont {Mei},
  \citenamefont {Mock}, \citenamefont {Moongweluwan}, \citenamefont {Morad},
  \citenamefont {Murphy}, \citenamefont {Nehrkorn}, \citenamefont {Nelson},
  \citenamefont {Neves}, \citenamefont {O'Sullivan}, \citenamefont
  {Oliver-Mallory}, \citenamefont {Palladino}, \citenamefont {Pease},
  \citenamefont {Phelps}, \citenamefont {Reichhart}, \citenamefont {Rhyne},
  \citenamefont {Shaw}, \citenamefont {Shutt}, \citenamefont {Silva},
  \citenamefont {Solmaz}, \citenamefont {Solovov}, \citenamefont {Sorensen},
  \citenamefont {Stephenson}, \citenamefont {Sumner}, \citenamefont {Szydagis},
  \citenamefont {Taylor}, \citenamefont {Taylor}, \citenamefont {Tennyson},
  \citenamefont {Terman}, \citenamefont {Tiedt}, \citenamefont {To},
  \citenamefont {Tripathi}, \citenamefont {Tvrznikova}, \citenamefont {Uvarov},
  \citenamefont {Verbus}, \citenamefont {Webb}, \citenamefont {White},
  \citenamefont {Whitis}, \citenamefont {Witherell}, \citenamefont {Wolfs},
  \citenamefont {Xu}, \citenamefont {Yazdani}, \citenamefont {Young},\ and\
  \citenamefont {Zhang}}]{Akerib2017}%
  \BibitemOpen
  \bibfield  {author} {\bibinfo {author} {\bibfnamefont {D.~S.}\ \bibnamefont
  {Akerib}}, \bibinfo {author} {\bibfnamefont {S.}~\bibnamefont {Alsum}},
  \bibinfo {author} {\bibfnamefont {H.~M.}\ \bibnamefont {Ara\'ujo}}, \bibinfo
  {author} {\bibfnamefont {X.}~\bibnamefont {Bai}}, \bibinfo {author}
  {\bibfnamefont {A.~J.}\ \bibnamefont {Bailey}}, \bibinfo {author}
  {\bibfnamefont {J.}~\bibnamefont {Balajthy}}, \bibinfo {author}
  {\bibfnamefont {P.}~\bibnamefont {Beltrame}}, \bibinfo {author}
  {\bibfnamefont {E.~P.}\ \bibnamefont {Bernard}}, \bibinfo {author}
  {\bibfnamefont {A.}~\bibnamefont {Bernstein}}, \bibinfo {author}
  {\bibfnamefont {T.~P.}\ \bibnamefont {Biesiadzinski}}, \bibinfo {author}
  {\bibfnamefont {E.~M.}\ \bibnamefont {Boulton}}, \bibinfo {author}
  {\bibfnamefont {R.}~\bibnamefont {Bramante}}, \bibinfo {author}
  {\bibfnamefont {P.}~\bibnamefont {Br\'as}}, \bibinfo {author} {\bibfnamefont
  {D.}~\bibnamefont {Byram}}, \bibinfo {author} {\bibfnamefont {S.~B.}\
  \bibnamefont {Cahn}}, \bibinfo {author} {\bibfnamefont {M.~C.}\ \bibnamefont
  {Carmona-Benitez}}, \bibinfo {author} {\bibfnamefont {C.}~\bibnamefont
  {Chan}}, \bibinfo {author} {\bibfnamefont {A.~A.}\ \bibnamefont {Chiller}},
  \bibinfo {author} {\bibfnamefont {C.}~\bibnamefont {Chiller}}, \bibinfo
  {author} {\bibfnamefont {A.}~\bibnamefont {Currie}}, \bibinfo {author}
  {\bibfnamefont {J.~E.}\ \bibnamefont {Cutter}}, \bibinfo {author}
  {\bibfnamefont {T.~J.~R.}\ \bibnamefont {Davison}}, \bibinfo {author}
  {\bibfnamefont {A.}~\bibnamefont {Dobi}}, \bibinfo {author} {\bibfnamefont
  {J.~E.~Y.}\ \bibnamefont {Dobson}}, \bibinfo {author} {\bibfnamefont
  {E.}~\bibnamefont {Druszkiewicz}}, \bibinfo {author} {\bibfnamefont {B.~N.}\
  \bibnamefont {Edwards}}, \bibinfo {author} {\bibfnamefont {C.~H.}\
  \bibnamefont {Faham}}, \bibinfo {author} {\bibfnamefont {S.}~\bibnamefont
  {Fiorucci}}, \bibinfo {author} {\bibfnamefont {R.~J.}\ \bibnamefont
  {Gaitskell}}, \bibinfo {author} {\bibfnamefont {V.~M.}\ \bibnamefont
  {Gehman}}, \bibinfo {author} {\bibfnamefont {C.}~\bibnamefont {Ghag}},
  \bibinfo {author} {\bibfnamefont {K.~R.}\ \bibnamefont {Gibson}}, \bibinfo
  {author} {\bibfnamefont {M.~G.~D.}\ \bibnamefont {Gilchriese}}, \bibinfo
  {author} {\bibfnamefont {C.~R.}\ \bibnamefont {Hall}}, \bibinfo {author}
  {\bibfnamefont {M.}~\bibnamefont {Hanhardt}}, \bibinfo {author}
  {\bibfnamefont {S.~J.}\ \bibnamefont {Haselschwardt}}, \bibinfo {author}
  {\bibfnamefont {S.~A.}\ \bibnamefont {Hertel}}, \bibinfo {author}
  {\bibfnamefont {D.~P.}\ \bibnamefont {Hogan}}, \bibinfo {author}
  {\bibfnamefont {M.}~\bibnamefont {Horn}}, \bibinfo {author} {\bibfnamefont
  {D.~Q.}\ \bibnamefont {Huang}}, \bibinfo {author} {\bibfnamefont {C.~M.}\
  \bibnamefont {Ignarra}}, \bibinfo {author} {\bibfnamefont {M.}~\bibnamefont
  {Ihm}}, \bibinfo {author} {\bibfnamefont {R.~G.}\ \bibnamefont {Jacobsen}},
  \bibinfo {author} {\bibfnamefont {W.}~\bibnamefont {Ji}}, \bibinfo {author}
  {\bibfnamefont {K.}~\bibnamefont {Kamdin}}, \bibinfo {author} {\bibfnamefont
  {K.}~\bibnamefont {Kazkaz}}, \bibinfo {author} {\bibfnamefont
  {D.}~\bibnamefont {Khaitan}}, \bibinfo {author} {\bibfnamefont
  {R.}~\bibnamefont {Knoche}}, \bibinfo {author} {\bibfnamefont {N.~A.}\
  \bibnamefont {Larsen}}, \bibinfo {author} {\bibfnamefont {C.}~\bibnamefont
  {Lee}}, \bibinfo {author} {\bibfnamefont {B.~G.}\ \bibnamefont {Lenardo}},
  \bibinfo {author} {\bibfnamefont {K.~T.}\ \bibnamefont {Lesko}}, \bibinfo
  {author} {\bibfnamefont {A.}~\bibnamefont {Lindote}}, \bibinfo {author}
  {\bibfnamefont {M.~I.}\ \bibnamefont {Lopes}}, \bibinfo {author}
  {\bibfnamefont {A.}~\bibnamefont {Manalaysay}}, \bibinfo {author}
  {\bibfnamefont {R.~L.}\ \bibnamefont {Mannino}}, \bibinfo {author}
  {\bibfnamefont {M.~F.}\ \bibnamefont {Marzioni}}, \bibinfo {author}
  {\bibfnamefont {D.~N.}\ \bibnamefont {McKinsey}}, \bibinfo {author}
  {\bibfnamefont {D.-M.}\ \bibnamefont {Mei}}, \bibinfo {author} {\bibfnamefont
  {J.}~\bibnamefont {Mock}}, \bibinfo {author} {\bibfnamefont {M.}~\bibnamefont
  {Moongweluwan}}, \bibinfo {author} {\bibfnamefont {J.~A.}\ \bibnamefont
  {Morad}}, \bibinfo {author} {\bibfnamefont {A.~S.~J.}\ \bibnamefont
  {Murphy}}, \bibinfo {author} {\bibfnamefont {C.}~\bibnamefont {Nehrkorn}},
  \bibinfo {author} {\bibfnamefont {H.~N.}\ \bibnamefont {Nelson}}, \bibinfo
  {author} {\bibfnamefont {F.}~\bibnamefont {Neves}}, \bibinfo {author}
  {\bibfnamefont {K.}~\bibnamefont {O'Sullivan}}, \bibinfo {author}
  {\bibfnamefont {K.~C.}\ \bibnamefont {Oliver-Mallory}}, \bibinfo {author}
  {\bibfnamefont {K.~J.}\ \bibnamefont {Palladino}}, \bibinfo {author}
  {\bibfnamefont {E.~K.}\ \bibnamefont {Pease}}, \bibinfo {author}
  {\bibfnamefont {P.}~\bibnamefont {Phelps}}, \bibinfo {author} {\bibfnamefont
  {L.}~\bibnamefont {Reichhart}}, \bibinfo {author} {\bibfnamefont
  {C.}~\bibnamefont {Rhyne}}, \bibinfo {author} {\bibfnamefont
  {S.}~\bibnamefont {Shaw}}, \bibinfo {author} {\bibfnamefont {T.~A.}\
  \bibnamefont {Shutt}}, \bibinfo {author} {\bibfnamefont {C.}~\bibnamefont
  {Silva}}, \bibinfo {author} {\bibfnamefont {M.}~\bibnamefont {Solmaz}},
  \bibinfo {author} {\bibfnamefont {V.~N.}\ \bibnamefont {Solovov}}, \bibinfo
  {author} {\bibfnamefont {P.}~\bibnamefont {Sorensen}}, \bibinfo {author}
  {\bibfnamefont {S.}~\bibnamefont {Stephenson}}, \bibinfo {author}
  {\bibfnamefont {T.~J.}\ \bibnamefont {Sumner}}, \bibinfo {author}
  {\bibfnamefont {M.}~\bibnamefont {Szydagis}}, \bibinfo {author}
  {\bibfnamefont {D.~J.}\ \bibnamefont {Taylor}}, \bibinfo {author}
  {\bibfnamefont {W.~C.}\ \bibnamefont {Taylor}}, \bibinfo {author}
  {\bibfnamefont {B.~P.}\ \bibnamefont {Tennyson}}, \bibinfo {author}
  {\bibfnamefont {P.~A.}\ \bibnamefont {Terman}}, \bibinfo {author}
  {\bibfnamefont {D.~R.}\ \bibnamefont {Tiedt}}, \bibinfo {author}
  {\bibfnamefont {W.~H.}\ \bibnamefont {To}}, \bibinfo {author} {\bibfnamefont
  {M.}~\bibnamefont {Tripathi}}, \bibinfo {author} {\bibfnamefont
  {L.}~\bibnamefont {Tvrznikova}}, \bibinfo {author} {\bibfnamefont
  {S.}~\bibnamefont {Uvarov}}, \bibinfo {author} {\bibfnamefont {J.~R.}\
  \bibnamefont {Verbus}}, \bibinfo {author} {\bibfnamefont {R.~C.}\
  \bibnamefont {Webb}}, \bibinfo {author} {\bibfnamefont {J.~T.}\ \bibnamefont
  {White}}, \bibinfo {author} {\bibfnamefont {T.~J.}\ \bibnamefont {Whitis}},
  \bibinfo {author} {\bibfnamefont {M.~S.}\ \bibnamefont {Witherell}}, \bibinfo
  {author} {\bibfnamefont {F.~L.~H.}\ \bibnamefont {Wolfs}}, \bibinfo {author}
  {\bibfnamefont {J.}~\bibnamefont {Xu}}, \bibinfo {author} {\bibfnamefont
  {K.}~\bibnamefont {Yazdani}}, \bibinfo {author} {\bibfnamefont {S.~K.}\
  \bibnamefont {Young}}, \ and\ \bibinfo {author} {\bibfnamefont
  {C.}~\bibnamefont {Zhang}} (\bibinfo {collaboration} {LUX Collaboration}),\
  }\href {\doibase 10.1103/PhysRevLett.118.021303} {\bibfield  {journal}
  {\bibinfo  {journal} {Phys. Rev. Lett.}\ }\textbf {\bibinfo {volume} {118}},\
  \bibinfo {pages} {021303} (\bibinfo {year} {2017})}\BibitemShut {NoStop}%
\bibitem [{\citenamefont {Cushman}\ \emph {et~al.}(2013)\citenamefont
  {Cushman}, \citenamefont {Galbiati}, \citenamefont {McKinsey}, \citenamefont
  {Robertson}, \citenamefont {Tait}, \citenamefont {Bauer}, \citenamefont
  {Borgland}, \citenamefont {Cabrera}, \citenamefont {Calaprice}, \citenamefont
  {Cooley}, \citenamefont {Empl}, \citenamefont {Essig}, \citenamefont
  {Figueroa-Feliciano}, \citenamefont {Gaitskell}, \citenamefont {Golwala},
  \citenamefont {Hall}, \citenamefont {Hill}, \citenamefont {Hime},
  \citenamefont {Hoppe}, \citenamefont {Hsu}, \citenamefont {Hungerford},
  \citenamefont {Jacobsen}, \citenamefont {Kelsey}, \citenamefont {Lang},
  \citenamefont {Lippincott}, \citenamefont {Loer}, \citenamefont {Luitz},
  \citenamefont {Mandic}, \citenamefont {Mardon}, \citenamefont {Maricic},
  \citenamefont {Maruyama}, \citenamefont {Mahapatra}, \citenamefont {Nelson},
  \citenamefont {Orrell}, \citenamefont {Palladino}, \citenamefont {Pantic},
  \citenamefont {Partridge}, \citenamefont {Ryd}, \citenamefont {Saab},
  \citenamefont {Sadoulet}, \citenamefont {Schnee}, \citenamefont {Shepherd},
  \citenamefont {Sonnenschein}, \citenamefont {Sorensen}, \citenamefont
  {Szydagis}, \citenamefont {Volansky}, \citenamefont {Witherell},
  \citenamefont {Wright},\ and\ \citenamefont {Zurek}}]{Cushman2013}%
  \BibitemOpen
  \bibfield  {author} {\bibinfo {author} {\bibfnamefont {P.}~\bibnamefont
  {Cushman}}, \bibinfo {author} {\bibfnamefont {C.}~\bibnamefont {Galbiati}},
  \bibinfo {author} {\bibfnamefont {D.~N.}\ \bibnamefont {McKinsey}}, \bibinfo
  {author} {\bibfnamefont {H.}~\bibnamefont {Robertson}}, \bibinfo {author}
  {\bibfnamefont {T.~M.~P.}\ \bibnamefont {Tait}}, \bibinfo {author}
  {\bibfnamefont {D.}~\bibnamefont {Bauer}}, \bibinfo {author} {\bibfnamefont
  {A.}~\bibnamefont {Borgland}}, \bibinfo {author} {\bibfnamefont
  {B.}~\bibnamefont {Cabrera}}, \bibinfo {author} {\bibfnamefont
  {F.}~\bibnamefont {Calaprice}}, \bibinfo {author} {\bibfnamefont
  {J.}~\bibnamefont {Cooley}}, \bibinfo {author} {\bibfnamefont
  {T.}~\bibnamefont {Empl}}, \bibinfo {author} {\bibfnamefont {R.}~\bibnamefont
  {Essig}}, \bibinfo {author} {\bibfnamefont {E.}~\bibnamefont
  {Figueroa-Feliciano}}, \bibinfo {author} {\bibfnamefont {R.}~\bibnamefont
  {Gaitskell}}, \bibinfo {author} {\bibfnamefont {S.}~\bibnamefont {Golwala}},
  \bibinfo {author} {\bibfnamefont {J.}~\bibnamefont {Hall}}, \bibinfo {author}
  {\bibfnamefont {R.}~\bibnamefont {Hill}}, \bibinfo {author} {\bibfnamefont
  {A.}~\bibnamefont {Hime}}, \bibinfo {author} {\bibfnamefont {E.}~\bibnamefont
  {Hoppe}}, \bibinfo {author} {\bibfnamefont {L.}~\bibnamefont {Hsu}}, \bibinfo
  {author} {\bibfnamefont {E.}~\bibnamefont {Hungerford}}, \bibinfo {author}
  {\bibfnamefont {R.}~\bibnamefont {Jacobsen}}, \bibinfo {author}
  {\bibfnamefont {M.}~\bibnamefont {Kelsey}}, \bibinfo {author} {\bibfnamefont
  {R.~F.}\ \bibnamefont {Lang}}, \bibinfo {author} {\bibfnamefont {W.~H.}\
  \bibnamefont {Lippincott}}, \bibinfo {author} {\bibfnamefont
  {B.}~\bibnamefont {Loer}}, \bibinfo {author} {\bibfnamefont {S.}~\bibnamefont
  {Luitz}}, \bibinfo {author} {\bibfnamefont {V.}~\bibnamefont {Mandic}},
  \bibinfo {author} {\bibfnamefont {J.}~\bibnamefont {Mardon}}, \bibinfo
  {author} {\bibfnamefont {J.}~\bibnamefont {Maricic}}, \bibinfo {author}
  {\bibfnamefont {R.}~\bibnamefont {Maruyama}}, \bibinfo {author}
  {\bibfnamefont {R.}~\bibnamefont {Mahapatra}}, \bibinfo {author}
  {\bibfnamefont {H.}~\bibnamefont {Nelson}}, \bibinfo {author} {\bibfnamefont
  {J.}~\bibnamefont {Orrell}}, \bibinfo {author} {\bibfnamefont
  {K.}~\bibnamefont {Palladino}}, \bibinfo {author} {\bibfnamefont
  {E.}~\bibnamefont {Pantic}}, \bibinfo {author} {\bibfnamefont
  {R.}~\bibnamefont {Partridge}}, \bibinfo {author} {\bibfnamefont
  {A.}~\bibnamefont {Ryd}}, \bibinfo {author} {\bibfnamefont {T.}~\bibnamefont
  {Saab}}, \bibinfo {author} {\bibfnamefont {B.}~\bibnamefont {Sadoulet}},
  \bibinfo {author} {\bibfnamefont {R.}~\bibnamefont {Schnee}}, \bibinfo
  {author} {\bibfnamefont {W.}~\bibnamefont {Shepherd}}, \bibinfo {author}
  {\bibfnamefont {A.}~\bibnamefont {Sonnenschein}}, \bibinfo {author}
  {\bibfnamefont {P.}~\bibnamefont {Sorensen}}, \bibinfo {author}
  {\bibfnamefont {M.}~\bibnamefont {Szydagis}}, \bibinfo {author}
  {\bibfnamefont {T.}~\bibnamefont {Volansky}}, \bibinfo {author}
  {\bibfnamefont {M.}~\bibnamefont {Witherell}}, \bibinfo {author}
  {\bibfnamefont {D.}~\bibnamefont {Wright}}, \ and\ \bibinfo {author}
  {\bibfnamefont {K.}~\bibnamefont {Zurek}},\ }\href@noop {} {\enquote
  {\bibinfo {title} {Snowmass cf1 summary: Wimp dark matter direct
  detection},}\ } (\bibinfo {year} {2013}),\ \Eprint
  {http://arxiv.org/abs/arXiv:1310.8327} {arXiv:1310.8327} \BibitemShut
  {NoStop}%
\bibitem [{\citenamefont {Strauss}\ \emph {et~al.}(2016)\citenamefont
  {Strauss}, \citenamefont {Angloher}, \citenamefont {Bento}, \citenamefont
  {Bucci}, \citenamefont {Canonica}, \citenamefont {Defay}, \citenamefont
  {Erb}, \citenamefont {Feilitzsch}, \citenamefont {Iachellini}, \citenamefont
  {Gorla} \emph {et~al.}}]{Strauss2016}%
  \BibitemOpen
  \bibfield  {author} {\bibinfo {author} {\bibfnamefont {R.}~\bibnamefont
  {Strauss}}, \bibinfo {author} {\bibfnamefont {G.}~\bibnamefont {Angloher}},
  \bibinfo {author} {\bibfnamefont {A.}~\bibnamefont {Bento}}, \bibinfo
  {author} {\bibfnamefont {C.}~\bibnamefont {Bucci}}, \bibinfo {author}
  {\bibfnamefont {L.}~\bibnamefont {Canonica}}, \bibinfo {author}
  {\bibfnamefont {X.}~\bibnamefont {Defay}}, \bibinfo {author} {\bibfnamefont
  {A.}~\bibnamefont {Erb}}, \bibinfo {author} {\bibfnamefont {F.~v.}\
  \bibnamefont {Feilitzsch}}, \bibinfo {author} {\bibfnamefont {N.~F.}\
  \bibnamefont {Iachellini}}, \bibinfo {author} {\bibfnamefont
  {P.}~\bibnamefont {Gorla}},  \emph {et~al.},\ }\href@noop {} {\bibfield
  {journal} {\bibinfo  {journal} {J. Low Temp. Phys.}\ }\textbf {\bibinfo
  {volume} {184}},\ \bibinfo {pages} {866} (\bibinfo {year}
  {2016})}\BibitemShut {NoStop}%
\bibitem [{\citenamefont {Angloher}\ \emph {et~al.}(2016)\citenamefont
  {Angloher}, \citenamefont {Bento}, \citenamefont {Bucci}, \citenamefont
  {Canonica}, \citenamefont {Defay}, \citenamefont {Erb}, \citenamefont {von
  Feilitzsch}, \citenamefont {Iachellini}, \citenamefont {Gorla}, \citenamefont
  {G{\"u}tlein} \emph {et~al.}}]{Angloher2016}%
  \BibitemOpen
  \bibfield  {author} {\bibinfo {author} {\bibfnamefont {G.}~\bibnamefont
  {Angloher}}, \bibinfo {author} {\bibfnamefont {A.}~\bibnamefont {Bento}},
  \bibinfo {author} {\bibfnamefont {C.}~\bibnamefont {Bucci}}, \bibinfo
  {author} {\bibfnamefont {L.}~\bibnamefont {Canonica}}, \bibinfo {author}
  {\bibfnamefont {X.}~\bibnamefont {Defay}}, \bibinfo {author} {\bibfnamefont
  {A.}~\bibnamefont {Erb}}, \bibinfo {author} {\bibfnamefont {F.}~\bibnamefont
  {von Feilitzsch}}, \bibinfo {author} {\bibfnamefont {N.~F.}\ \bibnamefont
  {Iachellini}}, \bibinfo {author} {\bibfnamefont {P.}~\bibnamefont {Gorla}},
  \bibinfo {author} {\bibfnamefont {A.}~\bibnamefont {G{\"u}tlein}},  \emph
  {et~al.},\ }\href@noop {} {\bibfield  {journal} {\bibinfo  {journal} {Eur.
  Phys. J. C}\ }\textbf {\bibinfo {volume} {76}},\ \bibinfo {pages} {1}
  (\bibinfo {year} {2016})}\BibitemShut {NoStop}%
\bibitem [{\citenamefont {Agnese}\ \emph {et~al.}(2014)\citenamefont {Agnese},
  \citenamefont {Anderson}, \citenamefont {Asai}, \citenamefont
  {Balakishiyeva}, \citenamefont {Thakur}, \citenamefont {Bauer}, \citenamefont
  {Beaty}, \citenamefont {Billard}, \citenamefont {Borgland}, \citenamefont
  {Bowles} \emph {et~al.}}]{Agnese2014}%
  \BibitemOpen
  \bibfield  {author} {\bibinfo {author} {\bibfnamefont {R.}~\bibnamefont
  {Agnese}}, \bibinfo {author} {\bibfnamefont {A.~J.}\ \bibnamefont
  {Anderson}}, \bibinfo {author} {\bibfnamefont {M.}~\bibnamefont {Asai}},
  \bibinfo {author} {\bibfnamefont {D.}~\bibnamefont {Balakishiyeva}}, \bibinfo
  {author} {\bibfnamefont {R.~B.}\ \bibnamefont {Thakur}}, \bibinfo {author}
  {\bibfnamefont {D.}~\bibnamefont {Bauer}}, \bibinfo {author} {\bibfnamefont
  {J.}~\bibnamefont {Beaty}}, \bibinfo {author} {\bibfnamefont
  {J.}~\bibnamefont {Billard}}, \bibinfo {author} {\bibfnamefont
  {A.}~\bibnamefont {Borgland}}, \bibinfo {author} {\bibfnamefont
  {M.}~\bibnamefont {Bowles}},  \emph {et~al.},\ }\href@noop {} {\bibfield
  {journal} {\bibinfo  {journal} {Phys. Rev. Lett.}\ }\textbf {\bibinfo
  {volume} {112}},\ \bibinfo {pages} {241302} (\bibinfo {year}
  {2014})}\BibitemShut {NoStop}%
\bibitem [{\citenamefont {Agnese}\ \emph {et~al.}(2015)\citenamefont {Agnese},
  \citenamefont {Anderson}, \citenamefont {Asai}, \citenamefont
  {Balakishiyeva}, \citenamefont {Barker}, \citenamefont {Thakur},
  \citenamefont {Bauer}, \citenamefont {Billard}, \citenamefont {Borgland},
  \citenamefont {Bowles} \emph {et~al.}}]{Agnese2015}%
  \BibitemOpen
  \bibfield  {author} {\bibinfo {author} {\bibfnamefont {R.}~\bibnamefont
  {Agnese}}, \bibinfo {author} {\bibfnamefont {A.}~\bibnamefont {Anderson}},
  \bibinfo {author} {\bibfnamefont {M.}~\bibnamefont {Asai}}, \bibinfo {author}
  {\bibfnamefont {D.}~\bibnamefont {Balakishiyeva}}, \bibinfo {author}
  {\bibfnamefont {D.}~\bibnamefont {Barker}}, \bibinfo {author} {\bibfnamefont
  {R.~B.}\ \bibnamefont {Thakur}}, \bibinfo {author} {\bibfnamefont
  {D.}~\bibnamefont {Bauer}}, \bibinfo {author} {\bibfnamefont
  {J.}~\bibnamefont {Billard}}, \bibinfo {author} {\bibfnamefont
  {A.}~\bibnamefont {Borgland}}, \bibinfo {author} {\bibfnamefont
  {M.}~\bibnamefont {Bowles}},  \emph {et~al.},\ }\href@noop {} {\bibfield
  {journal} {\bibinfo  {journal} {Phys. Rev. D}\ }\textbf {\bibinfo {volume}
  {92}},\ \bibinfo {pages} {072003} (\bibinfo {year} {2015})}\BibitemShut
  {NoStop}%
\bibitem [{\citenamefont {Armengaud}\ \emph {et~al.}()\citenamefont
  {Armengaud}, \citenamefont {Arnaud}, \citenamefont {Augier}, \citenamefont
  {Beno{\^\i}t}, \citenamefont {Berg{\'e}}, \citenamefont {Bergmann},
  \citenamefont {Billard}, \citenamefont {Bl{\"u}mer}, \citenamefont
  {De~Boissi{\`e}re}, \citenamefont {Bres} \emph {et~al.}}]{Armengaud2016}%
  \BibitemOpen
  \bibfield  {author} {\bibinfo {author} {\bibfnamefont {E.}~\bibnamefont
  {Armengaud}}, \bibinfo {author} {\bibfnamefont {Q.}~\bibnamefont {Arnaud}},
  \bibinfo {author} {\bibfnamefont {C.}~\bibnamefont {Augier}}, \bibinfo
  {author} {\bibfnamefont {A.}~\bibnamefont {Beno{\^\i}t}}, \bibinfo {author}
  {\bibfnamefont {L.}~\bibnamefont {Berg{\'e}}}, \bibinfo {author}
  {\bibfnamefont {T.}~\bibnamefont {Bergmann}}, \bibinfo {author}
  {\bibfnamefont {J.}~\bibnamefont {Billard}}, \bibinfo {author} {\bibfnamefont
  {J.}~\bibnamefont {Bl{\"u}mer}}, \bibinfo {author} {\bibfnamefont
  {T.}~\bibnamefont {De~Boissi{\`e}re}}, \bibinfo {author} {\bibfnamefont
  {G.}~\bibnamefont {Bres}},  \emph {et~al.},\ }\href@noop {} {\bibfield
  {journal} {\bibinfo  {journal} {J. Cosmol. Astropart. Phys.}\ }\textbf
  {\bibinfo {volume} {05}}}\BibitemShut {NoStop}%
\bibitem [{\citenamefont {Hehn}\ \emph {et~al.}(2016)\citenamefont {Hehn},
  \citenamefont {Armengaud}, \citenamefont {Arnaud}, \citenamefont {Augier},
  \citenamefont {Beno{\^\i}t}, \citenamefont {Berg{\'e}}, \citenamefont
  {Billard}, \citenamefont {Bl{\"u}mer}, \citenamefont {De~Boissi{\`e}re},
  \citenamefont {Broniatowski} \emph {et~al.}}]{Hehn2016}%
  \BibitemOpen
  \bibfield  {author} {\bibinfo {author} {\bibfnamefont {L.}~\bibnamefont
  {Hehn}}, \bibinfo {author} {\bibfnamefont {E.}~\bibnamefont {Armengaud}},
  \bibinfo {author} {\bibfnamefont {Q.}~\bibnamefont {Arnaud}}, \bibinfo
  {author} {\bibfnamefont {C.}~\bibnamefont {Augier}}, \bibinfo {author}
  {\bibfnamefont {A.}~\bibnamefont {Beno{\^\i}t}}, \bibinfo {author}
  {\bibfnamefont {L.}~\bibnamefont {Berg{\'e}}}, \bibinfo {author}
  {\bibfnamefont {J.}~\bibnamefont {Billard}}, \bibinfo {author} {\bibfnamefont
  {J.}~\bibnamefont {Bl{\"u}mer}}, \bibinfo {author} {\bibfnamefont
  {T.}~\bibnamefont {De~Boissi{\`e}re}}, \bibinfo {author} {\bibfnamefont
  {A.}~\bibnamefont {Broniatowski}},  \emph {et~al.},\ }\href@noop {}
  {\bibfield  {journal} {\bibinfo  {journal} {Eur. Phys. J. C}\ }\textbf
  {\bibinfo {volume} {76}},\ \bibinfo {pages} {548} (\bibinfo {year}
  {2016})}\BibitemShut {NoStop}%
\bibitem [{\citenamefont {Graham}\ \emph {et~al.}(2012)\citenamefont {Graham},
  \citenamefont {Kaplan}, \citenamefont {Rajendran},\ and\ \citenamefont
  {Walters}}]{Graham2012}%
  \BibitemOpen
  \bibfield  {author} {\bibinfo {author} {\bibfnamefont {P.~W.}\ \bibnamefont
  {Graham}}, \bibinfo {author} {\bibfnamefont {D.~E.}\ \bibnamefont {Kaplan}},
  \bibinfo {author} {\bibfnamefont {S.}~\bibnamefont {Rajendran}}, \ and\
  \bibinfo {author} {\bibfnamefont {M.~T.}\ \bibnamefont {Walters}},\
  }\href@noop {} {\bibfield  {journal} {\bibinfo  {journal} {Phys. Dark
  Universe.}\ }\textbf {\bibinfo {volume} {1}},\ \bibinfo {pages} {32}
  (\bibinfo {year} {2012})}\BibitemShut {NoStop}%
\bibitem [{\citenamefont {Essig}\ \emph {et~al.}(2012)\citenamefont {Essig},
  \citenamefont {Mardon},\ and\ \citenamefont {Volansky}}]{Essig2012}%
  \BibitemOpen
  \bibfield  {author} {\bibinfo {author} {\bibfnamefont {R.}~\bibnamefont
  {Essig}}, \bibinfo {author} {\bibfnamefont {J.}~\bibnamefont {Mardon}}, \
  and\ \bibinfo {author} {\bibfnamefont {T.}~\bibnamefont {Volansky}},\
  }\href@noop {} {\bibfield  {journal} {\bibinfo  {journal} {Phys. Rev. D}\
  }\textbf {\bibinfo {volume} {85}},\ \bibinfo {pages} {076007} (\bibinfo
  {year} {2012})}\BibitemShut {NoStop}%
\bibitem [{\citenamefont {Hochberg}\ \emph {et~al.}(2016)\citenamefont
  {Hochberg}, \citenamefont {Zhao},\ and\ \citenamefont
  {Zurek}}]{Hochberg2016}%
  \BibitemOpen
  \bibfield  {author} {\bibinfo {author} {\bibfnamefont {Y.}~\bibnamefont
  {Hochberg}}, \bibinfo {author} {\bibfnamefont {Y.}~\bibnamefont {Zhao}}, \
  and\ \bibinfo {author} {\bibfnamefont {K.~M.}\ \bibnamefont {Zurek}},\ }\href
  {\doibase 10.1103/PhysRevLett.116.011301} {\bibfield  {journal} {\bibinfo
  {journal} {Phys. Rev. Lett.}\ }\textbf {\bibinfo {volume} {116}},\ \bibinfo
  {pages} {011301} (\bibinfo {year} {2016})}\BibitemShut {NoStop}%
\bibitem [{\citenamefont {Guo}\ and\ \citenamefont {McKinsey}(2013)}]{Guo2013}%
  \BibitemOpen
  \bibfield  {author} {\bibinfo {author} {\bibfnamefont {W.}~\bibnamefont
  {Guo}}\ and\ \bibinfo {author} {\bibfnamefont {D.~N.}\ \bibnamefont
  {McKinsey}},\ }\href {\doibase 10.1103/PhysRevD.87.115001} {\bibfield
  {journal} {\bibinfo  {journal} {Phys. Rev. D}\ }\textbf {\bibinfo {volume}
  {87}},\ \bibinfo {pages} {115001} (\bibinfo {year} {2013})}\BibitemShut
  {NoStop}%
\bibitem [{\citenamefont {Ito}\ and\ \citenamefont {Seidel}(2013)}]{Ito2013}%
  \BibitemOpen
  \bibfield  {author} {\bibinfo {author} {\bibfnamefont {T.~M.}\ \bibnamefont
  {Ito}}\ and\ \bibinfo {author} {\bibfnamefont {G.~M.}\ \bibnamefont
  {Seidel}},\ }\href@noop {} {\bibfield  {journal} {\bibinfo  {journal} {Phys.
  Rev. C}\ }\textbf {\bibinfo {volume} {88}},\ \bibinfo {pages} {025805}
  (\bibinfo {year} {2013})}\BibitemShut {NoStop}%
\bibitem [{\citenamefont {Schutz}\ and\ \citenamefont
  {Zurek}(2016)}]{Schutz2016}%
  \BibitemOpen
  \bibfield  {author} {\bibinfo {author} {\bibfnamefont {K.}~\bibnamefont
  {Schutz}}\ and\ \bibinfo {author} {\bibfnamefont {K.~M.}\ \bibnamefont
  {Zurek}},\ }\href {\doibase 10.1103/PhysRevLett.117.121302} {\bibfield
  {journal} {\bibinfo  {journal} {Phys. Rev. Lett.}\ }\textbf {\bibinfo
  {volume} {117}},\ \bibinfo {pages} {121302} (\bibinfo {year}
  {2016})}\BibitemShut {NoStop}%
\bibitem [{\citenamefont {Lanou}\ \emph {et~al.}(1987)\citenamefont {Lanou},
  \citenamefont {Maris},\ and\ \citenamefont {Seidel}}]{Lanou1987}%
  \BibitemOpen
  \bibfield  {author} {\bibinfo {author} {\bibfnamefont {R.~E.}\ \bibnamefont
  {Lanou}}, \bibinfo {author} {\bibfnamefont {H.~J.}\ \bibnamefont {Maris}}, \
  and\ \bibinfo {author} {\bibfnamefont {G.~M.}\ \bibnamefont {Seidel}},\
  }\href {\doibase 10.1103/PhysRevLett.58.2498} {\bibfield  {journal} {\bibinfo
   {journal} {Phys. Rev. Lett.}\ }\textbf {\bibinfo {volume} {58}},\ \bibinfo
  {pages} {2498} (\bibinfo {year} {1987})}\BibitemShut {NoStop}%
\bibitem [{\citenamefont {Bandler}\ \emph {et~al.}(1992)\citenamefont
  {Bandler}, \citenamefont {Lanou}, \citenamefont {Maris}, \citenamefont
  {More}, \citenamefont {Porter}, \citenamefont {Seidel},\ and\ \citenamefont
  {Torii}}]{Bandler1992}%
  \BibitemOpen
  \bibfield  {author} {\bibinfo {author} {\bibfnamefont {S.~R.}\ \bibnamefont
  {Bandler}}, \bibinfo {author} {\bibfnamefont {R.~E.}\ \bibnamefont {Lanou}},
  \bibinfo {author} {\bibfnamefont {H.~J.}\ \bibnamefont {Maris}}, \bibinfo
  {author} {\bibfnamefont {T.}~\bibnamefont {More}}, \bibinfo {author}
  {\bibfnamefont {F.~S.}\ \bibnamefont {Porter}}, \bibinfo {author}
  {\bibfnamefont {G.~M.}\ \bibnamefont {Seidel}}, \ and\ \bibinfo {author}
  {\bibfnamefont {R.~H.}\ \bibnamefont {Torii}},\ }\href {\doibase
  10.1103/PhysRevLett.68.2429} {\bibfield  {journal} {\bibinfo  {journal}
  {Phys. Rev. Lett.}\ }\textbf {\bibinfo {volume} {68}},\ \bibinfo {pages}
  {2429} (\bibinfo {year} {1992})}\BibitemShut {NoStop}%
\bibitem [{\citenamefont {Bandler}\ \emph {et~al.}(1995)\citenamefont
  {Bandler}, \citenamefont {Brou\"er}, \citenamefont {Enss}, \citenamefont
  {Lanou}, \citenamefont {Maris}, \citenamefont {More}, \citenamefont
  {Porter},\ and\ \citenamefont {Seidel}}]{Bandler1995}%
  \BibitemOpen
  \bibfield  {author} {\bibinfo {author} {\bibfnamefont {S.~R.}\ \bibnamefont
  {Bandler}}, \bibinfo {author} {\bibfnamefont {S.~M.}\ \bibnamefont
  {Brou\"er}}, \bibinfo {author} {\bibfnamefont {C.}~\bibnamefont {Enss}},
  \bibinfo {author} {\bibfnamefont {R.~E.}\ \bibnamefont {Lanou}}, \bibinfo
  {author} {\bibfnamefont {H.~J.}\ \bibnamefont {Maris}}, \bibinfo {author}
  {\bibfnamefont {T.}~\bibnamefont {More}}, \bibinfo {author} {\bibfnamefont
  {F.~S.}\ \bibnamefont {Porter}}, \ and\ \bibinfo {author} {\bibfnamefont
  {G.~M.}\ \bibnamefont {Seidel}},\ }\href {\doibase
  10.1103/PhysRevLett.74.3169} {\bibfield  {journal} {\bibinfo  {journal}
  {Phys. Rev. Lett.}\ }\textbf {\bibinfo {volume} {74}},\ \bibinfo {pages}
  {3169} (\bibinfo {year} {1995})}\BibitemShut {NoStop}%
\bibitem [{\citenamefont {Enss}\ \emph {et~al.}(1994)\citenamefont {Enss},
  \citenamefont {Bandler}, \citenamefont {Lanou}, \citenamefont {Maris},
  \citenamefont {More}, \citenamefont {Porter},\ and\ \citenamefont
  {Seidel}}]{Enss1994}%
  \BibitemOpen
  \bibfield  {author} {\bibinfo {author} {\bibfnamefont {C.}~\bibnamefont
  {Enss}}, \bibinfo {author} {\bibfnamefont {S.}~\bibnamefont {Bandler}},
  \bibinfo {author} {\bibfnamefont {R.}~\bibnamefont {Lanou}}, \bibinfo
  {author} {\bibfnamefont {H.}~\bibnamefont {Maris}}, \bibinfo {author}
  {\bibfnamefont {T.}~\bibnamefont {More}}, \bibinfo {author} {\bibfnamefont
  {F.}~\bibnamefont {Porter}}, \ and\ \bibinfo {author} {\bibfnamefont {G.~M.}\
  \bibnamefont {Seidel}},\ }\href@noop {} {\bibfield  {journal} {\bibinfo
  {journal} {Physica B Condens Matter.}\ }\textbf {\bibinfo {volume} {194}},\
  \bibinfo {pages} {515} (\bibinfo {year} {1994})}\BibitemShut {NoStop}%
\bibitem [{\citenamefont {Adams}\ \emph {et~al.}(1998)\citenamefont {Adams},
  \citenamefont {Kim}, \citenamefont {Lanou}, \citenamefont {Maris},\ and\
  \citenamefont {Seidel}}]{Adams1998}%
  \BibitemOpen
  \bibfield  {author} {\bibinfo {author} {\bibfnamefont {J.~S.}\ \bibnamefont
  {Adams}}, \bibinfo {author} {\bibfnamefont {Y.~H.}\ \bibnamefont {Kim}},
  \bibinfo {author} {\bibfnamefont {R.~E.}\ \bibnamefont {Lanou}}, \bibinfo
  {author} {\bibfnamefont {H.~J.}\ \bibnamefont {Maris}}, \ and\ \bibinfo
  {author} {\bibfnamefont {G.~M.}\ \bibnamefont {Seidel}},\ }\href@noop {}
  {\bibfield  {journal} {\bibinfo  {journal} {J. Low Temp. Phys.}\ }\textbf
  {\bibinfo {volume} {113}},\ \bibinfo {pages} {1121} (\bibinfo {year}
  {1998})}\BibitemShut {NoStop}%
\bibitem [{\citenamefont {Adams}\ \emph {et~al.}(2000)\citenamefont {Adams},
  \citenamefont {Fleischmann}, \citenamefont {Huang}, \citenamefont {Kim},
  \citenamefont {Lanou}, \citenamefont {Maris},\ and\ \citenamefont
  {Seidel}}]{Adams2000}%
  \BibitemOpen
  \bibfield  {author} {\bibinfo {author} {\bibfnamefont {J.}~\bibnamefont
  {Adams}}, \bibinfo {author} {\bibfnamefont {A.}~\bibnamefont {Fleischmann}},
  \bibinfo {author} {\bibfnamefont {Y.}~\bibnamefont {Huang}}, \bibinfo
  {author} {\bibfnamefont {Y.}~\bibnamefont {Kim}}, \bibinfo {author}
  {\bibfnamefont {R.}~\bibnamefont {Lanou}}, \bibinfo {author} {\bibfnamefont
  {H.}~\bibnamefont {Maris}}, \ and\ \bibinfo {author} {\bibfnamefont {G.~M.}\
  \bibnamefont {Seidel}},\ }\href@noop {} {\bibfield  {journal} {\bibinfo
  {journal} {Nucl. Instr. Meth. Phys. Res. A}\ }\textbf {\bibinfo {volume}
  {444}},\ \bibinfo {pages} {51} (\bibinfo {year} {2000})}\BibitemShut
  {NoStop}%
\bibitem [{\citenamefont {Maris}(1977)}]{Maris1977}%
  \BibitemOpen
  \bibfield  {author} {\bibinfo {author} {\bibfnamefont {H.~J.}\ \bibnamefont
  {Maris}},\ }\href {\doibase 10.1103/RevModPhys.49.341} {\bibfield  {journal}
  {\bibinfo  {journal} {Rev. Mod. Phys.}\ }\textbf {\bibinfo {volume} {49}},\
  \bibinfo {pages} {341} (\bibinfo {year} {1977})}\BibitemShut {NoStop}%
\bibitem [{Not()}]{NoteLosses}%
  \BibitemOpen
  \href@noop {} {}\bibinfo {note} {Except possibly some losses at the walls of
  the helium container.}\BibitemShut {Stop}%
\bibitem [{\citenamefont {Brown}\ and\ \citenamefont
  {Wyatt}(1990)}]{Brown1990}%
  \BibitemOpen
  \bibfield  {author} {\bibinfo {author} {\bibfnamefont {M.}~\bibnamefont
  {Brown}}\ and\ \bibinfo {author} {\bibfnamefont {A.}~\bibnamefont {Wyatt}},\
  }\href@noop {} {\bibfield  {journal} {\bibinfo  {journal} {J. Phys. Condens.
  Matter}\ }\textbf {\bibinfo {volume} {2}},\ \bibinfo {pages} {5025} (\bibinfo
  {year} {1990})}\BibitemShut {NoStop}%
\bibitem [{\citenamefont {Wyatt}(1992)}]{Wyatt1992}%
  \BibitemOpen
  \bibfield  {author} {\bibinfo {author} {\bibfnamefont {A.}~\bibnamefont
  {Wyatt}},\ }\href@noop {} {\bibfield  {journal} {\bibinfo  {journal} {J. Low
  Temp. Phys.}\ }\textbf {\bibinfo {volume} {87}},\ \bibinfo {pages} {453}
  (\bibinfo {year} {1992})}\BibitemShut {NoStop}%
\bibitem [{\citenamefont {Balibar}(2016)}]{Balibar2016}%
  \BibitemOpen
  \bibfield  {author} {\bibinfo {author} {\bibfnamefont {S.}~\bibnamefont
  {Balibar}},\ }\href@noop {} {\bibfield  {journal} {\bibinfo  {journal} {J.
  Low Temp. Phys.}\ }\textbf {\bibinfo {volume} {185}},\ \bibinfo {pages} {209}
  (\bibinfo {year} {2016})}\BibitemShut {NoStop}%
\bibitem [{\citenamefont {Torii}\ \emph {et~al.}(1992)\citenamefont {Torii},
  \citenamefont {Bandler}, \citenamefont {More}, \citenamefont {Porter},
  \citenamefont {Lanou}, \citenamefont {Maris},\ and\ \citenamefont
  {Seidel}}]{Torii1992}%
  \BibitemOpen
  \bibfield  {author} {\bibinfo {author} {\bibfnamefont {R.}~\bibnamefont
  {Torii}}, \bibinfo {author} {\bibfnamefont {S.}~\bibnamefont {Bandler}},
  \bibinfo {author} {\bibfnamefont {T.}~\bibnamefont {More}}, \bibinfo {author}
  {\bibfnamefont {F.}~\bibnamefont {Porter}}, \bibinfo {author} {\bibfnamefont
  {R.}~\bibnamefont {Lanou}}, \bibinfo {author} {\bibfnamefont
  {H.}~\bibnamefont {Maris}}, \ and\ \bibinfo {author} {\bibfnamefont {G.~M.}\
  \bibnamefont {Seidel}},\ }\href@noop {} {\bibfield  {journal} {\bibinfo
  {journal} {Rev. Sci. Instrum.}\ }\textbf {\bibinfo {volume} {63}},\ \bibinfo
  {pages} {230} (\bibinfo {year} {1992})}\BibitemShut {NoStop}%
\bibitem [{\citenamefont {M{\"u}ller}(1951)}]{Muller1951}%
  \BibitemOpen
  \bibfield  {author} {\bibinfo {author} {\bibfnamefont {E.~W.}\ \bibnamefont
  {M{\"u}ller}},\ }\href@noop {} {\bibfield  {journal} {\bibinfo  {journal} {Z.
  Phys.}\ }\textbf {\bibinfo {volume} {131}},\ \bibinfo {pages} {136} (\bibinfo
  {year} {1951})}\BibitemShut {NoStop}%
\bibitem [{\citenamefont {M{\"u}ller}\ and\ \citenamefont
  {Bahadur}(1956)}]{Muller1956}%
  \BibitemOpen
  \bibfield  {author} {\bibinfo {author} {\bibfnamefont {E.~W.}\ \bibnamefont
  {M{\"u}ller}}\ and\ \bibinfo {author} {\bibfnamefont {K.}~\bibnamefont
  {Bahadur}},\ }\href@noop {} {\bibfield  {journal} {\bibinfo  {journal} {Phys.
  Rev.}\ }\textbf {\bibinfo {volume} {102}},\ \bibinfo {pages} {624} (\bibinfo
  {year} {1956})}\BibitemShut {NoStop}%
\bibitem [{\citenamefont {Gomer}(1994)}]{Gomer1994}%
  \BibitemOpen
  \bibfield  {author} {\bibinfo {author} {\bibfnamefont {R.}~\bibnamefont
  {Gomer}},\ }\href@noop {} {\bibfield  {journal} {\bibinfo  {journal} {Surf.
  Sci.}\ }\textbf {\bibinfo {volume} {299}},\ \bibinfo {pages} {129} (\bibinfo
  {year} {1994})}\BibitemShut {NoStop}%
\bibitem [{\citenamefont {O'Donnell}\ \emph {et~al.}(2010)\citenamefont
  {O'Donnell}, \citenamefont {Fahy}, \citenamefont {Thomsen}, \citenamefont
  {O'Connor},\ and\ \citenamefont {Dastoor}}]{ODonnell2010}%
  \BibitemOpen
  \bibfield  {author} {\bibinfo {author} {\bibfnamefont {K.}~\bibnamefont
  {O'Donnell}}, \bibinfo {author} {\bibfnamefont {A.}~\bibnamefont {Fahy}},
  \bibinfo {author} {\bibfnamefont {L.}~\bibnamefont {Thomsen}}, \bibinfo
  {author} {\bibfnamefont {D.}~\bibnamefont {O'Connor}}, \ and\ \bibinfo
  {author} {\bibfnamefont {P.}~\bibnamefont {Dastoor}},\ }\href@noop {}
  {\bibfield  {journal} {\bibinfo  {journal} {Meas. Sci. Technol.}\ }\textbf
  {\bibinfo {volume} {22}},\ \bibinfo {pages} {015901} (\bibinfo {year}
  {2010})}\BibitemShut {NoStop}%
\bibitem [{\citenamefont {Johnston~Jr}\ and\ \citenamefont
  {King}(1966)}]{Johnston1966}%
  \BibitemOpen
  \bibfield  {author} {\bibinfo {author} {\bibfnamefont {W.~D.}\ \bibnamefont
  {Johnston~Jr}}\ and\ \bibinfo {author} {\bibfnamefont {J.~G.}\ \bibnamefont
  {King}},\ }\href@noop {} {\bibfield  {journal} {\bibinfo  {journal} {Rev.
  Sci. Instrum.}\ }\textbf {\bibinfo {volume} {37}},\ \bibinfo {pages} {475}
  (\bibinfo {year} {1966})}\BibitemShut {NoStop}%
\bibitem [{\citenamefont {Goodman}(1980)}]{Goodman1980}%
  \BibitemOpen
  \bibfield  {author} {\bibinfo {author} {\bibfnamefont {F.~O.}\ \bibnamefont
  {Goodman}},\ }\href@noop {} {\bibfield  {journal} {\bibinfo  {journal} {J.
  Phys. Chem.}\ }\textbf {\bibinfo {volume} {84}},\ \bibinfo {pages} {1431}
  (\bibinfo {year} {1980})}\BibitemShut {NoStop}%
\bibitem [{\citenamefont {Southon}\ and\ \citenamefont
  {Brandon}(1963)}]{Southon1963}%
  \BibitemOpen
  \bibfield  {author} {\bibinfo {author} {\bibfnamefont {M.~J.}\ \bibnamefont
  {Southon}}\ and\ \bibinfo {author} {\bibfnamefont {D.~G.}\ \bibnamefont
  {Brandon}},\ }\href@noop {} {\bibfield  {journal} {\bibinfo  {journal} {Phil.
  Mag.}\ }\textbf {\bibinfo {volume} {8}},\ \bibinfo {pages} {579} (\bibinfo
  {year} {1963})}\BibitemShut {NoStop}%
\bibitem [{\citenamefont {B{\"o}rret}\ \emph {et~al.}(1990)\citenamefont
  {B{\"o}rret}, \citenamefont {B{\"o}hringer},\ and\ \citenamefont
  {Kalbitzer}}]{Borret1990}%
  \BibitemOpen
  \bibfield  {author} {\bibinfo {author} {\bibfnamefont {R.}~\bibnamefont
  {B{\"o}rret}}, \bibinfo {author} {\bibfnamefont {K.}~\bibnamefont
  {B{\"o}hringer}}, \ and\ \bibinfo {author} {\bibfnamefont {S.}~\bibnamefont
  {Kalbitzer}},\ }\href@noop {} {\bibfield  {journal} {\bibinfo  {journal} {J.
  Phys. D: Appl. Phys.}\ }\textbf {\bibinfo {volume} {23}},\ \bibinfo {pages}
  {1271} (\bibinfo {year} {1990})}\BibitemShut {NoStop}%
\bibitem [{\citenamefont {Piskur}\ \emph {et~al.}(2008)\citenamefont {Piskur},
  \citenamefont {Borg}, \citenamefont {Stupnik}, \citenamefont {Leisch},
  \citenamefont {Ernst},\ and\ \citenamefont {Holst}}]{Piskur2008}%
  \BibitemOpen
  \bibfield  {author} {\bibinfo {author} {\bibfnamefont {J.}~\bibnamefont
  {Piskur}}, \bibinfo {author} {\bibfnamefont {L.}~\bibnamefont {Borg}},
  \bibinfo {author} {\bibfnamefont {A.}~\bibnamefont {Stupnik}}, \bibinfo
  {author} {\bibfnamefont {M.}~\bibnamefont {Leisch}}, \bibinfo {author}
  {\bibfnamefont {W.}~\bibnamefont {Ernst}}, \ and\ \bibinfo {author}
  {\bibfnamefont {B.}~\bibnamefont {Holst}},\ }\href@noop {} {\bibfield
  {journal} {\bibinfo  {journal} {Appl. Surf. Sci.}\ }\textbf {\bibinfo
  {volume} {254}},\ \bibinfo {pages} {4365} (\bibinfo {year}
  {2008})}\BibitemShut {NoStop}%
\bibitem [{\citenamefont {Halpern}\ and\ \citenamefont
  {Gomer}(1969)}]{Halpern1969}%
  \BibitemOpen
  \bibfield  {author} {\bibinfo {author} {\bibfnamefont {B.}~\bibnamefont
  {Halpern}}\ and\ \bibinfo {author} {\bibfnamefont {R.}~\bibnamefont
  {Gomer}},\ }\href@noop {} {\bibfield  {journal} {\bibinfo  {journal} {J.
  Chem. Phys.}\ }\textbf {\bibinfo {volume} {51}},\ \bibinfo {pages} {5709}
  (\bibinfo {year} {1969})}\BibitemShut {NoStop}%
\bibitem [{\citenamefont {Edgcombe}\ and\ \citenamefont
  {Valdre}(2001)}]{Edgcombe2001}%
  \BibitemOpen
  \bibfield  {author} {\bibinfo {author} {\bibfnamefont {C.~J.}\ \bibnamefont
  {Edgcombe}}\ and\ \bibinfo {author} {\bibfnamefont {U.}~\bibnamefont
  {Valdre}},\ }\href@noop {} {\bibfield  {journal} {\bibinfo  {journal} {Solid
  State Electron.}\ }\textbf {\bibinfo {volume} {45}},\ \bibinfo {pages} {857}
  (\bibinfo {year} {2001})}\BibitemShut {NoStop}%
\bibitem [{\citenamefont {Read}\ and\ \citenamefont
  {Bowring}(2004)}]{Read2004}%
  \BibitemOpen
  \bibfield  {author} {\bibinfo {author} {\bibfnamefont {F.}~\bibnamefont
  {Read}}\ and\ \bibinfo {author} {\bibfnamefont {N.}~\bibnamefont {Bowring}},\
  }\href@noop {} {\bibfield  {journal} {\bibinfo  {journal} {Nucl. Instr. Meth.
  Phys. Res. A}\ }\textbf {\bibinfo {volume} {519}},\ \bibinfo {pages} {305}
  (\bibinfo {year} {2004})}\BibitemShut {NoStop}%
\bibitem [{\citenamefont {Johnson}\ \emph {et~al.}(2012)\citenamefont
  {Johnson}, \citenamefont {Schwoebel}, \citenamefont {Holland}, \citenamefont
  {Resnick}, \citenamefont {Hertz},\ and\ \citenamefont
  {Chichester}}]{Johnson2012}%
  \BibitemOpen
  \bibfield  {author} {\bibinfo {author} {\bibfnamefont {B.~B.}\ \bibnamefont
  {Johnson}}, \bibinfo {author} {\bibfnamefont {P.}~\bibnamefont {Schwoebel}},
  \bibinfo {author} {\bibfnamefont {C.}~\bibnamefont {Holland}}, \bibinfo
  {author} {\bibfnamefont {P.}~\bibnamefont {Resnick}}, \bibinfo {author}
  {\bibfnamefont {K.}~\bibnamefont {Hertz}}, \ and\ \bibinfo {author}
  {\bibfnamefont {D.}~\bibnamefont {Chichester}},\ }\href@noop {} {\bibfield
  {journal} {\bibinfo  {journal} {Nucl. Instr. Meth. Phys. Res. A}\ }\textbf
  {\bibinfo {volume} {663}},\ \bibinfo {pages} {64} (\bibinfo {year}
  {2012})}\BibitemShut {NoStop}%
\bibitem [{\citenamefont {Modi}\ \emph {et~al.}(2003)\citenamefont {Modi},
  \citenamefont {Koratkar}, \citenamefont {Lass}, \citenamefont {Wei},\ and\
  \citenamefont {Ajayan}}]{Modi2003}%
  \BibitemOpen
  \bibfield  {author} {\bibinfo {author} {\bibfnamefont {A.}~\bibnamefont
  {Modi}}, \bibinfo {author} {\bibfnamefont {N.}~\bibnamefont {Koratkar}},
  \bibinfo {author} {\bibfnamefont {E.}~\bibnamefont {Lass}}, \bibinfo {author}
  {\bibfnamefont {B.}~\bibnamefont {Wei}}, \ and\ \bibinfo {author}
  {\bibfnamefont {P.~M.}\ \bibnamefont {Ajayan}},\ }\href@noop {} {\bibfield
  {journal} {\bibinfo  {journal} {Nature}\ }\textbf {\bibinfo {volume} {424}},\
  \bibinfo {pages} {171} (\bibinfo {year} {2003})}\BibitemShut {NoStop}%
\bibitem [{\citenamefont {Gesemann}\ \emph {et~al.}(2011)\citenamefont
  {Gesemann}, \citenamefont {Wehrspohn}, \citenamefont {Hackner},\ and\
  \citenamefont {Muller}}]{Gesemann2011}%
  \BibitemOpen
  \bibfield  {author} {\bibinfo {author} {\bibfnamefont {B.}~\bibnamefont
  {Gesemann}}, \bibinfo {author} {\bibfnamefont {R.}~\bibnamefont {Wehrspohn}},
  \bibinfo {author} {\bibfnamefont {A.}~\bibnamefont {Hackner}}, \ and\
  \bibinfo {author} {\bibfnamefont {G.}~\bibnamefont {Muller}},\ }\href@noop {}
  {\bibfield  {journal} {\bibinfo  {journal} {IEEE T. NANOTECHNOL.}\ }\textbf
  {\bibinfo {volume} {10}},\ \bibinfo {pages} {50} (\bibinfo {year}
  {2011})}\BibitemShut {NoStop}%
\bibitem [{\citenamefont {Liu}\ \emph {et~al.}(2013)\citenamefont {Liu},
  \citenamefont {Ma}, \citenamefont {Shi}, \citenamefont {Zhou}, \citenamefont
  {Gong}, \citenamefont {Lei}, \citenamefont {Yang}, \citenamefont {Zhang},
  \citenamefont {Yu}, \citenamefont {Hackenberg} \emph {et~al.}}]{Liu2013}%
  \BibitemOpen
  \bibfield  {author} {\bibinfo {author} {\bibfnamefont {Z.}~\bibnamefont
  {Liu}}, \bibinfo {author} {\bibfnamefont {L.}~\bibnamefont {Ma}}, \bibinfo
  {author} {\bibfnamefont {G.}~\bibnamefont {Shi}}, \bibinfo {author}
  {\bibfnamefont {W.}~\bibnamefont {Zhou}}, \bibinfo {author} {\bibfnamefont
  {Y.}~\bibnamefont {Gong}}, \bibinfo {author} {\bibfnamefont {S.}~\bibnamefont
  {Lei}}, \bibinfo {author} {\bibfnamefont {X.}~\bibnamefont {Yang}}, \bibinfo
  {author} {\bibfnamefont {J.}~\bibnamefont {Zhang}}, \bibinfo {author}
  {\bibfnamefont {J.}~\bibnamefont {Yu}}, \bibinfo {author} {\bibfnamefont
  {K.~P.}\ \bibnamefont {Hackenberg}},  \emph {et~al.},\ }\href@noop {}
  {\bibfield  {journal} {\bibinfo  {journal} {Nat. Nanotechnol.}\ }\textbf
  {\bibinfo {volume} {8}},\ \bibinfo {pages} {119} (\bibinfo {year}
  {2013})}\BibitemShut {NoStop}%
\bibitem [{\citenamefont {Spitsina}\ and\ \citenamefont
  {Kahrizi}(2016)}]{Spitsina2016}%
  \BibitemOpen
  \bibfield  {author} {\bibinfo {author} {\bibfnamefont {S.}~\bibnamefont
  {Spitsina}}\ and\ \bibinfo {author} {\bibfnamefont {M.}~\bibnamefont
  {Kahrizi}},\ }\href@noop {} {\bibfield  {journal} {\bibinfo  {journal}
  {Sensor Mater.}\ }\textbf {\bibinfo {volume} {28}},\ \bibinfo {pages} {43}
  (\bibinfo {year} {2016})}\BibitemShut {NoStop}%
\bibitem [{\citenamefont {Mohammadpour}\ \emph {et~al.}(2014)\citenamefont
  {Mohammadpour}, \citenamefont {Ahmadvand} \emph {et~al.}}]{Mohammadpour2014}%
  \BibitemOpen
  \bibfield  {author} {\bibinfo {author} {\bibfnamefont {R.}~\bibnamefont
  {Mohammadpour}}, \bibinfo {author} {\bibfnamefont {H.}~\bibnamefont
  {Ahmadvand}},  \emph {et~al.},\ }\href@noop {} {\bibfield  {journal}
  {\bibinfo  {journal} {Sens. Actuators A Phys.}\ }\textbf {\bibinfo {volume}
  {216}},\ \bibinfo {pages} {202} (\bibinfo {year} {2014})}\BibitemShut
  {NoStop}%
\bibitem [{\citenamefont {Ohno}\ \emph {et~al.}(1978)\citenamefont {Ohno},
  \citenamefont {Nakamura},\ and\ \citenamefont {Kuroda}}]{Ohno1978}%
  \BibitemOpen
  \bibfield  {author} {\bibinfo {author} {\bibfnamefont {Y.}~\bibnamefont
  {Ohno}}, \bibinfo {author} {\bibfnamefont {S.}~\bibnamefont {Nakamura}}, \
  and\ \bibinfo {author} {\bibfnamefont {T.}~\bibnamefont {Kuroda}},\
  }\href@noop {} {\bibfield  {journal} {\bibinfo  {journal} {Jpn. J. Appl.
  Phys.}\ }\textbf {\bibinfo {volume} {17}},\ \bibinfo {pages} {2013} (\bibinfo
  {year} {1978})}\BibitemShut {NoStop}%
\bibitem [{\citenamefont {Tsong}(1979)}]{Tsong1979}%
  \BibitemOpen
  \bibfield  {author} {\bibinfo {author} {\bibfnamefont {T.~T.}\ \bibnamefont
  {Tsong}},\ }\href@noop {} {\bibfield  {journal} {\bibinfo  {journal} {Surf.
  Sci.}\ }\textbf {\bibinfo {volume} {81}},\ \bibinfo {pages} {28} (\bibinfo
  {year} {1979})}\BibitemShut {NoStop}%
\bibitem [{\citenamefont {Suzuki}\ \emph {et~al.}(2001)\citenamefont {Suzuki},
  \citenamefont {Nakanishi}, \citenamefont {Okumi}, \citenamefont {Gotou},
  \citenamefont {Togawa}, \citenamefont {Furuta}, \citenamefont {Wada},
  \citenamefont {Nishitani}, \citenamefont {Yamamoto}, \citenamefont {Watanabe}
  \emph {et~al.}}]{Suzuki2001}%
  \BibitemOpen
  \bibfield  {author} {\bibinfo {author} {\bibfnamefont {C.}~\bibnamefont
  {Suzuki}}, \bibinfo {author} {\bibfnamefont {T.}~\bibnamefont {Nakanishi}},
  \bibinfo {author} {\bibfnamefont {S.}~\bibnamefont {Okumi}}, \bibinfo
  {author} {\bibfnamefont {T.}~\bibnamefont {Gotou}}, \bibinfo {author}
  {\bibfnamefont {K.}~\bibnamefont {Togawa}}, \bibinfo {author} {\bibfnamefont
  {F.}~\bibnamefont {Furuta}}, \bibinfo {author} {\bibfnamefont
  {K.}~\bibnamefont {Wada}}, \bibinfo {author} {\bibfnamefont {T.}~\bibnamefont
  {Nishitani}}, \bibinfo {author} {\bibfnamefont {M.}~\bibnamefont {Yamamoto}},
  \bibinfo {author} {\bibfnamefont {J.}~\bibnamefont {Watanabe}},  \emph
  {et~al.},\ }\href@noop {} {\bibfield  {journal} {\bibinfo  {journal} {Nucl.
  Instr. Meth. Phys. Res. A}\ }\textbf {\bibinfo {volume} {462}},\ \bibinfo
  {pages} {337} (\bibinfo {year} {2001})}\BibitemShut {NoStop}%
\bibitem [{\citenamefont {Lepetit}\ \emph {et~al.}(2016)\citenamefont
  {Lepetit}, \citenamefont {Lemoine},\ and\ \citenamefont
  {M{\'a}rquez-Mijares}}]{Lepetit2016}%
  \BibitemOpen
  \bibfield  {author} {\bibinfo {author} {\bibfnamefont {B.}~\bibnamefont
  {Lepetit}}, \bibinfo {author} {\bibfnamefont {D.}~\bibnamefont {Lemoine}}, \
  and\ \bibinfo {author} {\bibfnamefont {M.}~\bibnamefont
  {M{\'a}rquez-Mijares}},\ }\href@noop {} {\bibfield  {journal} {\bibinfo
  {journal} {J. Appl. Phys.}\ }\textbf {\bibinfo {volume} {120}},\ \bibinfo
  {pages} {085105} (\bibinfo {year} {2016})}\BibitemShut {NoStop}%
\bibitem [{\citenamefont {Bandler}(1995)}]{BandlerPhD1995}%
  \BibitemOpen
  \bibfield  {author} {\bibinfo {author} {\bibfnamefont {S.~R.}\ \bibnamefont
  {Bandler}},\ }\emph {\bibinfo {title} {Ph.D. Thesis}},\ \href@noop {} {Ph.D.
  thesis},\ \bibinfo  {school} {Brown University} (\bibinfo {year}
  {1995})\BibitemShut {NoStop}%
\bibitem [{\citenamefont {Guilleumas}\ \emph {et~al.}(1998)\citenamefont
  {Guilleumas}, \citenamefont {Dalfovo}, \citenamefont {Oberosler},
  \citenamefont {Pitaevskii},\ and\ \citenamefont
  {Stringari}}]{Guilleumas1998}%
  \BibitemOpen
  \bibfield  {author} {\bibinfo {author} {\bibfnamefont {M.}~\bibnamefont
  {Guilleumas}}, \bibinfo {author} {\bibfnamefont {F.}~\bibnamefont {Dalfovo}},
  \bibinfo {author} {\bibfnamefont {I.}~\bibnamefont {Oberosler}}, \bibinfo
  {author} {\bibfnamefont {L.}~\bibnamefont {Pitaevskii}}, \ and\ \bibinfo
  {author} {\bibfnamefont {S.}~\bibnamefont {Stringari}},\ }\href@noop {}
  {\bibfield  {journal} {\bibinfo  {journal} {J. Low Temp. Phys.}\ }\textbf
  {\bibinfo {volume} {110}},\ \bibinfo {pages} {449} (\bibinfo {year}
  {1998})}\BibitemShut {NoStop}%
\bibitem [{\citenamefont {Wyatt}\ \emph {et~al.}(1976)\citenamefont {Wyatt},
  \citenamefont {Page},\ and\ \citenamefont {Sherlock}}]{Wyatt1976}%
  \BibitemOpen
  \bibfield  {author} {\bibinfo {author} {\bibfnamefont {A.~F.~G.}\
  \bibnamefont {Wyatt}}, \bibinfo {author} {\bibfnamefont {G.~J.}\ \bibnamefont
  {Page}}, \ and\ \bibinfo {author} {\bibfnamefont {R.~A.}\ \bibnamefont
  {Sherlock}},\ }\href@noop {} {\bibfield  {journal} {\bibinfo  {journal}
  {Phys. Rev. Lett.}\ }\textbf {\bibinfo {volume} {36}},\ \bibinfo {pages}
  {1184} (\bibinfo {year} {1976})}\BibitemShut {NoStop}%
\bibitem [{\citenamefont {Khalatnikov}(1965)}]{Khalatnikov1965}%
  \BibitemOpen
  \bibfield  {author} {\bibinfo {author} {\bibfnamefont {I.~M.}\ \bibnamefont
  {Khalatnikov}},\ }\href {https://cds.cern.ch/record/106134} {\emph {\bibinfo
  {title} {An introduction to the theory of superfluidity}}},\ Frontiers in
  Physics\ (\bibinfo  {publisher} {Benjamin},\ \bibinfo {address} {New York,
  NY},\ \bibinfo {year} {1965})\ \bibinfo {note} {trans. from the
  Russian}\BibitemShut {NoStop}%
\bibitem [{\citenamefont {Sinvani}\ \emph {et~al.}(1983)\citenamefont
  {Sinvani}, \citenamefont {Taborek},\ and\ \citenamefont
  {Goodstein}}]{Sinvani1983}%
  \BibitemOpen
  \bibfield  {author} {\bibinfo {author} {\bibfnamefont {M.}~\bibnamefont
  {Sinvani}}, \bibinfo {author} {\bibfnamefont {P.}~\bibnamefont {Taborek}}, \
  and\ \bibinfo {author} {\bibfnamefont {D.}~\bibnamefont {Goodstein}},\
  }\href@noop {} {\bibfield  {journal} {\bibinfo  {journal} {Physics Letters
  A}\ }\textbf {\bibinfo {volume} {95}},\ \bibinfo {pages} {59} (\bibinfo
  {year} {1983})}\BibitemShut {NoStop}%
\bibitem [{\citenamefont {More}\ \emph {et~al.}(1996)\citenamefont {More},
  \citenamefont {Adams}, \citenamefont {Bandler}, \citenamefont {Brou\"er},
  \citenamefont {Lanou}, \citenamefont {Maris},\ and\ \citenamefont
  {Seidel}}]{More1996}%
  \BibitemOpen
  \bibfield  {author} {\bibinfo {author} {\bibfnamefont {T.}~\bibnamefont
  {More}}, \bibinfo {author} {\bibfnamefont {J.~S.}\ \bibnamefont {Adams}},
  \bibinfo {author} {\bibfnamefont {S.~R.}\ \bibnamefont {Bandler}}, \bibinfo
  {author} {\bibfnamefont {S.~M.}\ \bibnamefont {Brou\"er}}, \bibinfo {author}
  {\bibfnamefont {R.~E.}\ \bibnamefont {Lanou}}, \bibinfo {author}
  {\bibfnamefont {H.~J.}\ \bibnamefont {Maris}}, \ and\ \bibinfo {author}
  {\bibfnamefont {G.~M.}\ \bibnamefont {Seidel}},\ }\href {\doibase
  10.1103/PhysRevB.54.534} {\bibfield  {journal} {\bibinfo  {journal} {Phys.
  Rev. B}\ }\textbf {\bibinfo {volume} {54}},\ \bibinfo {pages} {534} (\bibinfo
  {year} {1996})}\BibitemShut {NoStop}%
\bibitem [{\citenamefont {Van~Cleve}\ \emph {et~al.}(2008)\citenamefont
  {Van~Cleve}, \citenamefont {Taborek},\ and\ \citenamefont
  {Rutledge}}]{Van2008}%
  \BibitemOpen
  \bibfield  {author} {\bibinfo {author} {\bibfnamefont {E.}~\bibnamefont
  {Van~Cleve}}, \bibinfo {author} {\bibfnamefont {P.}~\bibnamefont {Taborek}},
  \ and\ \bibinfo {author} {\bibfnamefont {J.~E.}\ \bibnamefont {Rutledge}},\
  }\href@noop {} {\bibfield  {journal} {\bibinfo  {journal} {J. Low Temp.
  Phys.}\ }\textbf {\bibinfo {volume} {150}},\ \bibinfo {pages} {1} (\bibinfo
  {year} {2008})}\BibitemShut {NoStop}%
\bibitem [{\citenamefont {Rutledge}\ and\ \citenamefont
  {Taborek}(1992)}]{Rutledge1992}%
  \BibitemOpen
  \bibfield  {author} {\bibinfo {author} {\bibfnamefont {J.~E.}\ \bibnamefont
  {Rutledge}}\ and\ \bibinfo {author} {\bibfnamefont {P.}~\bibnamefont
  {Taborek}},\ }\href@noop {} {\bibfield  {journal} {\bibinfo  {journal} {Phys.
  Rev. Lett.}\ }\textbf {\bibinfo {volume} {69}},\ \bibinfo {pages} {937}
  (\bibinfo {year} {1992})}\BibitemShut {NoStop}%
\bibitem [{\citenamefont {Taborek}\ and\ \citenamefont
  {Rutledge}(1992)}]{Taborek1992}%
  \BibitemOpen
  \bibfield  {author} {\bibinfo {author} {\bibfnamefont {P.}~\bibnamefont
  {Taborek}}\ and\ \bibinfo {author} {\bibfnamefont {J.~E.}\ \bibnamefont
  {Rutledge}},\ }\href@noop {} {\bibfield  {journal} {\bibinfo  {journal}
  {Phys. Rev. Lett.}\ }\textbf {\bibinfo {volume} {68}},\ \bibinfo {pages}
  {2184} (\bibinfo {year} {1992})}\BibitemShut {NoStop}%
\bibitem [{\citenamefont {Gatica}\ and\ \citenamefont
  {Cole}(2009)}]{Gatica2009}%
  \BibitemOpen
  \bibfield  {author} {\bibinfo {author} {\bibfnamefont {S.~M.}\ \bibnamefont
  {Gatica}}\ and\ \bibinfo {author} {\bibfnamefont {M.~W.}\ \bibnamefont
  {Cole}},\ }\href@noop {} {\bibfield  {journal} {\bibinfo  {journal} {J. Low
  Temp. Phys.}\ }\textbf {\bibinfo {volume} {157}},\ \bibinfo {pages} {111}
  (\bibinfo {year} {2009})}\BibitemShut {NoStop}%
\bibitem [{\citenamefont {Ross}\ \emph {et~al.}(1995)\citenamefont {Ross},
  \citenamefont {Taborek},\ and\ \citenamefont {Rutledge}}]{Ross1995}%
  \BibitemOpen
  \bibfield  {author} {\bibinfo {author} {\bibfnamefont {D.}~\bibnamefont
  {Ross}}, \bibinfo {author} {\bibfnamefont {P.}~\bibnamefont {Taborek}}, \
  and\ \bibinfo {author} {\bibfnamefont {J.~E.}\ \bibnamefont {Rutledge}},\
  }\href@noop {} {\bibfield  {journal} {\bibinfo  {journal} {Phys. Rev. Lett.}\
  }\textbf {\bibinfo {volume} {74}},\ \bibinfo {pages} {4483} (\bibinfo {year}
  {1995})}\BibitemShut {NoStop}%
\bibitem [{\citenamefont {Nacher}\ and\ \citenamefont
  {Dupont-Roc}(1991)}]{Nacher1991}%
  \BibitemOpen
  \bibfield  {author} {\bibinfo {author} {\bibfnamefont {P.~J.}\ \bibnamefont
  {Nacher}}\ and\ \bibinfo {author} {\bibfnamefont {J.}~\bibnamefont
  {Dupont-Roc}},\ }\href@noop {} {\bibfield  {journal} {\bibinfo  {journal}
  {Phys. Rev. Lett.}\ }\textbf {\bibinfo {volume} {67}},\ \bibinfo {pages}
  {2966} (\bibinfo {year} {1991})}\BibitemShut {NoStop}%
\end{thebibliography}%

\end{document}